\documentclass[twocolumn,aps,showpacs,prb,tightenlines,amsmath,amssymb]{revtex4}
\usepackage{graphicx}
\usepackage{amssymb}
\usepackage{slashed}
\usepackage{dcolumn}
\usepackage{amsmath}
\usepackage{bm}% bold math
\usepackage{colordvi}
\usepackage{mathrsfs}
\makeatletter

\newcommand{\Rmnum}[1]{\expandafter\@slowromancap\romannumeral #1@}
\makeatother

\begin{document}

\title{A theory of coupled dual dynamics of macroscopic phase coherence and microscopic electronic fluids: effect of dephasing on cuprate superconductivity}   

\author{F. Yang}
\email{yfgq@mail.ustc.edu.cn.}

\affiliation{Hefei National Laboratory for Physical Sciences at
Microscale, Department of Physics, and CAS Key Laboratory of Strongly-Coupled
Quantum Matter Physics, University of Science and Technology of China, Hefei,
Anhui, 230026, China}

\author{M. W. Wu}
\email{mwwu@ustc.edu.cn.}

\affiliation{Hefei National Laboratory for Physical Sciences at
Microscale, Department of Physics, and CAS Key Laboratory of Strongly-Coupled
Quantum Matter Physics, University of Science and Technology of China, Hefei,
Anhui, 230026, China}

\date{\today}

\begin{abstract}
  By using the gauge-invariant kinetic equation approach [Yang and Wu, Phys. Rev. B {\bf 98}, 094507 (2018); {\bf 100}, 104513 (2019)], we construct the coupled dual dynamics of macroscopic phase coherence and microscopic electronic fluids in cuprate superconductors. We prove that the developed dual dynamics provides an efficient and simplified approach to formulate the dephasing process of macroscopic superconducting phase coherence, as well as its
  influence on microscopic electronic fluids (including gap, densities of superfluid and normal fluid, and in particular, the transport property to determine superconducting transition temperature $T_c$). We then present theoretical description of the preformed Cooper pairs in pseudogap state. The key origin of pseudogap state comes from the quantum effect of disorder, which excites the macroscopic inhomogeneous phase fluctuation through Josephson effect. Influenced by this phase fluctuation, 
  there exist normal fluid and viscous superfluid below $T_c$ in cuprate superconductors, in addition to conventional non-viscous superfluid. The normal fluid always emerges around nodal points even at zero temperature, whereas the viscous superfluid emerges due to the friction between superfluid and normal fluid. Particularly, the non-viscous superfluid gets suppressed when the phase fluctuation is enhanced by increasing temperature, until vanishes at $T_c$. Then, the system enters the pseudogap state, showing the nonzero resistivity as well as the finite gap from the viscous superfluid. By further increasing temperature to $T^{\rm os}$, the viscous superfluid and hence gap vanish. An experimental scheme to distinguish the densities of normal fluid as well as viscous and non-viscous superfluids is proposed. Finally, this theory is also applied to low-dimensional disordered $s$-wave superconductors.

\end{abstract}


%74.72.Kf Pseudogap regime
%74.62.En Effects of disorder
%74.72.-h Cuprate superconductors
%74.40.+k Fluctuations (noise, chaos, nonequilibrium superconductivity, localization, etc.)
%74.78.�6�0w Superconducting films and low-dimensional structures

\maketitle 

\section{Introduction}

Within the framework of Bardeen, Cooper and Schrieffer (BCS) theory of conventional superconductivity\cite{BCS}, the emergence of the superconducting order parameter by forming Cooper pairs occurs upon cooling at the critical temperature $T_c$, below which superconductivity (zero-resistance) phenomenon occurs. The high-$T_c$ superconducting materials, like cuprates first discovered in 1986\cite{cuprate1,cuprate2}, are beyond the BCS mechanism\cite{cuprate3,cuprate4,cuprate5}. Specifically, it is established that the superconducting order parameter shows up as an energy gap in spectroscopic probes of quasiparticle energy spectrum. Whereas in cuprate superconductors, detected most directly by angle-resolved photo-emission spectroscopy\cite{ARPES1,ARPES2,ARPES3} and scanning tunneling microscope\cite{STM1,STM2,STM3,STM4}, a finite normal-state gap (pseudogap) with $d$-wave symmetry opens below a temperature $T^*$, far above the superconducting transition temperature $T_c$. 

The origin of the pseudogap and its relationship to superconductivity have attracted extensive experimental and theoretical interest. Transport properties like electrical resistivity\cite{resistivity1,resistivity2}, Hall conductivity\cite{Hall1,Hall2} and Nernst coefficient\cite{Nernst1,Nernst2,Nernst3,Nernst4,Nernst5,Nernst6} that are remarkably affected by the opening of the pseudogap have been used to detect $T^*$. While among various theoretical models, significant superconducting phase fluctuation is suggested to be a possible attribution\cite{phase1,phase2,phase3}. Specifically, the generation of the superconducting order parameter breaks the continuous $U(1)$ symmetry spontaneously\cite{gi0}, and then, according to Goldstone theorem\cite{Gm1,Gm2}, a collective gapless Bosonic excitation that describes the phase fluctuation of the order parameter emerges\cite{gi0,AK,Ba0,Am0}. This excitation in conventional bulk superconductors is inactive\cite{AK,Ba0,Am0}, since its original low-energy spectrum is lifted to high-frequency plasma energy by Anderson-Higgs mechanism\cite{AHM}, due to the coupling to longitudinal electromagnetic field and hence long-range Coulomb interaction\cite{AK,Ba0,Am0}. Whereas in high-$T_c$ superconductors, an active phase fluctuation becomes inevitable because of the low-dimensional layered structure\cite{cuprate4,cuprate5}. Then, in one view, the pseudogap state is a incoherent precursor of the superconducting state\cite{phase2,PCP1,PCP2,PCP3,PCP4,PCP5}, reflecting a state of preformed Cooper pairs without the phase coherence necessary to achieve superconductivity. The superconducting transition then occurs upon cooling below a lower temperature where the long-range phase coherence is established. $T_c$ is therefore determined by the onset of the phase coherence rather than the formation of the Cooper pairs\cite{phase2,PCP1,PCP2,PCP3,PCP4,PCP5}. 

The existence of the preformed Cooper pairing above $T_c$ received a number of experimental supports by various approaches like diamagnetism probe\cite{TM1,TM2,TM3,TM4}, specific heat\cite{SH} and paraconductivity\cite{PC1} measurements, Nernst effect\cite{NF1}, ultrafast pump-probe spectroscopies\cite{PPS1,PPS2} and angle-resolved photo-emission\cite{AF1,AF2,AF3,AF4} as well as detection of optical conductivity in infrared\cite{Infrad1,Infrad2,Infrad3}, microwave\cite{MS1,MS2,MS3} and terahertz\cite{THZ1,THZ2,THZ3,THZ4} frequency regimes. The most convincing evidence so far comes from a recent observation of the collective Higgs mode above $T_c$\cite{DHM1,DHM2,DHM3}, since this excitation describes the amplitude fluctuation of the order parameter\cite{Am0} and thus directly reflects the existence of the pairing. Nevertheless, most of these experiments\cite{TM2,TM3,TM4,SH,PC1,NF1,AF1,AF2,AF3,AF4,PPS1,PPS2,Infrad1,Infrad2,Infrad3,MS1,MS2,MS3,THZ1,THZ2,THZ3,THZ4,DHM3} realized that the regime of the significant superconducting phase fluctuation is in fact a relatively narrow one that tracks $T_c$, whereas the upper onset temperature of this regime $T^{\rm os}$ lies well below the pseudogap temperature $T^*$ in underdoped regime and tends to coincide with $T^*$ in overdoped regime\cite{TM4,SH,PC1,NF1,AF4,MS1,MS2,MS3,THZ1,THZ2,THZ3,THZ4,DHM3}. Then, in an alternative view, the pseudogap state between $T^{\rm os}$ and $T^{*}$ in underdoped regime, exhibiting various intertwined orders (nematicity, charge-density-wave and spin-density-wave orders), possibly represents another state of matter that competes/couples with superconductivity\cite{STM1,AF4,NF1}. Whereas the precursor of the superconducting state with significant phase fluctuation (incoherent preformed Cooper pair) actually begins upon cooling at $T^{\rm os}$.

The phenomenological preformed Cooper-pair model with significant phase fluctuation is now widely accepted from experimental findings, but its microscopic theoretical description is not yet developed. Interestingly, from earlier thermodynamic\cite{NSB1} and recent optical\cite{NSB2} measurements, substantial fraction of the uncondensed normal state, which exhibits $T$-linear specific heat and Drude optical conductivity, persists down to temperatures far below $T_c$. This suggests that the phase fluctuation exists not only in the pseudogap state but also in the superconducting one. Therefore, to achieve superconductivity, the phase coherence must exceed a specific {\em nonzero} threshold. The theory of determining this threshold and in particular, $T_c$, requires coupled dual dynamics with different scales, i.e., in different Hilbert spaces: macroscopic phase-coherence dynamics and microscopic electronic fluid (including superfluid and normal fluid\cite{SNT}) dynamics. In the early-stage works, without distinguishing the amplitude and phase fluctuations, Ussishkin {\em et al.}\cite{GT1,GT2} applied the Gaussian approximation\cite{GT} to calculate the contribution from the fluctuations of order parameter to Nernst signal. After that, within the path-integral method to study thermodynamics, Curty and Beck\cite{MT1} treated separately amplitude and phase fluctuations, and used the Monte Carlo procedure and Wolff algorithm for simulations, respectively. But in both approaches, the origin of the fluctuations is unclear. Recently, Li {\em et al.}\cite{disorder} suggested that in cuprate superconductors, the disorder effect at low temperature plays the crucial role in determining the amplitude and phase fluctuations of the order parameter. By using tight-binding model with the random on-site potential (Anderson disorder)\cite{disorder}, they numerically obtained granular superconducting islands to explain the substantial fraction of the normal state at $T=0$, and then, suggested that the strong phase fluctuation can emerge in regions with small gap as a consequence. But this stationary-state calculation actually does not take account of the phase-coherence dynamics seriously, and is hard to extend for finite temperatures. Most importantly, the microscopic electronic fluid dynamics is decoupled and overlooked in all approaches above, inhibiting the deep insight into the key issue, separation between $T_c$ and $T^{\rm os}$.

Actually, a microscopic gauge-invariant kinetic equation (GIKE) approach has been developed in conventional $s$-wave superconductors\cite{GIKE1,GIKE2,GIKE3}. This approach, as analytically demonstrated, not only involves both superfluid and normal-fluid dynamics\cite{GIKE1}, but also is capable of formulating both phase and amplitude fluctuations of the order parameter (i.e., Nambu-Goldstone and Higgs modes)\cite{GIKE2}. The complete microscopic scattering is also constructed in GIKE\cite{GIKE1,GIKE3}. Very recently, this approach has also been extended into the $d$-wave superconductors for studying the Higgs modes\cite{GIKE4}. It is therefore natural to further use this approach to study the phase-coherence dynamics and its influence on electronic fluid dynamics in cuprate superconductors, and then, elucidate the fundamental nature of the incoherent preformed Cooper pairs.

In this work, by using the GIKE approach\cite{GIKE1,GIKE2,GIKE3,GIKE4}, we construct the coupled dual dynamics of macroscopic phase coherence and microscopic electronic fluids (consisting of normal fluid and superfluid) in cuprate superconductors. Then, both the dephasing process of macroscopic superconducting phase coherence from long range to short range with the increase of temperature, and the influence of this dephasing on microscopic electronic fluids (including gap, densities of superfluid and normal fluid, and in particular, the transport property to determine superconducting transition temperature $T_c$) can be formulated. 

Specifically, to develop the macroscopic phase-coherence dynamics, the equation of motion of the superconducting phase fluctuation, in which both disorder and long-range Coulomb interaction effects are considered, is derived analytically. We show that differing from the conventional bulk superconductors with inactive phase fluctuation due to Anderson-Higgs mechanism\cite{AHM,AK,Am0,Ba0}, the phase fluctuation in cuprate superconductors retains gapless energy spectrum after considering the long-range Coulomb interaction, thanks to the layered structures\cite{cuprate4,cuprate5}, and hence, is active. We also find that the superfluid density determines the superconducting phase stiffness in the phase-coherence dynamics, in consistency with the previous understanding in the literature\cite{phase1,phase2,phase3}. The phase-coherence dynamics is therefore influenced by electronic fluids. The derived microscopic electronic-fluid dynamics includes two parts: the anomalous correlation to reflect pairing (i.e., distinguish superfluid and normal fluid\cite{SNT}) and determine gap and superfluid density; the microscopic scattering of the electronic fluids that is essential for studying the transport property and hence superconductivity. It is established that the spatial fluctuation of superconducting phase can generate a superconducting momentum ${\bf p}_s$ \cite{gi0,Ba0,G1}, which drives the Doppler shift ${\bf v}_{\bf k}\!\cdot\!{\bf p}_s$ in quasiparticle energy spectra\cite{FF4,FF5,FF6,GIKE1}, with ${\bf v}_{\bf k}$ being the group velocity. Then, both anomalous correlation and microscopic scattering in the electronic-fluid dynamics are affected by the superconducting phase fluctuation through this Doppler shift.  Therefore, the macroscopic phase-coherence dynamics and microscopic electronic-fluid dynamics are mutually coupled.

It is noted that $T^{\rm os}$ and $T_c$ are determined by critical temperatures where the gap and resistivity vanish, respectively. Based on the developed dual dynamics, we present theoretical descriptions of the separation between $T_c$ and $T^{\rm os}$ as well as the emerged normal fluid in superconducting state (below $T_c$). The quantum effect of disorder, which excites a macroscopic inhomogeneous phase fluctuation through the Josephson effect\cite{Josephson}, provides the key origin. This excited phase fluctuation drives the Doppler shift ${\bf v}_{\bf k}\cdot{\bf p}_s$ mentioned above. Following the idea of Fulde-Ferrell-Larkin-Ovchinnikov (FFLO) state in conventional superconductors, the anomalous correlation vanishes in region where $|{\bf v}_{\bf k}\cdot{\bf p}_s|>\Delta_{\bf k}$ with $\Delta_{\bf k}$ being the superconducting gap\cite{FF1,FF2,FF7,FF8,FF9,GIKE1}. Particles then no longer participate in the pairing in this region and behave like the normal ones, leading to the emergence of normal fluid\cite{FF1,GIKE1}. Interestingly, with the phase fluctuation, we find that the condition of the unpairing region is always satisfied around nodal points in $d$-wave superconductors. One therefore always finds a nonzero fraction of the normal fluid even at low temperature, in consistency with the experimentally observed substantial fraction of normal state at low temperature\cite{NSB1,NSB2}.

Particles in regions with nonzero anomalous correlation contribute to the superconducting gap as superfluid. Moreover, we find that there exists the scattering between particles in pairing and unpairing regions in $d$-wave superconductors, and this scattering behaves like the friction between superfluid and normal fluid. We prove analytically that due to this friction, part of superfluid becomes viscous with nonzero momentum-relaxation rate. Consequently, in addition to conventional non-viscous superfluid, there also exist normal fluid and viscous superfluid at small phase fluctuation in cuprate superconductors, similar to the three-fluid model proposed in our previous work\cite{GIKE1} in conventional superconductors which is caused by external electromagnetic field. A scheme to detect distinguish these three electronic fluids in cuprate superconductors is then proposed. 

Particularly, as shown in Ref.~\onlinecite{GIKE1}, the increase of Doppler shift leads to the increases of normal fluid and viscous superfluid but the shrinkage of the non-viscous superfluid. When the non-viscous superfluid vanishes at large enough Doppler shift, one can find an exotic state with only normal fluid and viscous superfluid left, showing the nonzero resistivity as well as the finite gap from the viscous superfluid. Following the same idea, we demonstrate that by increasing the temperature in $d$-wave cuprate superconductors, the suppressed superconducting gap and hence superfluid density weaken the phase stiffness, enhancing the phase fluctuation and hence Doppler shift. Once the phase fluctuation (i.e., temperature) exceeds the critical point, the non-viscous superfluid vanishes, leaving only normal fluid and viscous superfluid. The system then enters the pseudogap state with nonzero resistivity and finite gap due to the significant phase fluctuation. It is noted that in this circumstance, the viscous superfluid matches the description of the incoherent preformed Cooper pairs, as they both contribute to gap but experience the scattering. Whereas the existing normal fluid in our description implies the existence of normal particles in pseudogap state, which has been overlooked in previous preformed Cooper-pair model to describe pseudogap state\cite{phase2,PCP1,PCP2,PCP3,PCP4,PCP5}. With a further increase of temperature in pseudogap state, the viscous superfluid starts to shrink, until vanishes at $T^{\rm os}$ where the gap is eventually destroyed.

To confirm the derivation from the GIKE approach, we also apply the standard path-integral approach to recover the equation of motion of the phase fluctuation as well as anomalous correlation, gap equation and superfluid density in the presence of the superconducting momentum, except the microscopic scattering of the electronic fluids which is hard to handle within the path-integral approach. A self-consistent numerical simulation by applying Anderson disorder is also addressed, to verify our theoretical description. Finally, we show that the developed dual dynamics can also be applied similarly to the low-dimensional disordered $s$-wave superconductors.

\section{Coupled dual dynamics}
\label{MDDsec}

In this section, from the rigorous analytic derivation within GIKE approach (refer to Sec.~\ref{sec-F}), we summarize the simplified results to present the coupled dual dynamics of macroscopic phase coherence and microscopic electronic fluids for a generalized order parameter:
\begin{equation}
\Delta(x,x')=\sum_{\bf k}e^{i{\bf k}\cdot({\bf r-r'})}[\Delta_{{\bf k}}+\delta\Delta_{\bf k}(R)]e^{i\delta\theta(R)}.  
\end{equation}
Here, $x=(x_0,{\bf r})$ denotes the space-time vector; $R={(x+x')}/{2}=(t,{\bf R})$ stands for the center-of-mass coordinate; $\Delta_{{\bf k}}$ denotes the equilibrium-state gap, independent of the center-of-mass coordinate due to the translational symmetry; $\delta\theta(R)$ and $\delta\Delta_{\bf k}(R)$ represent the phase and amplitude fluctuations, respectively.

Specifically, the macroscopic phase-coherence dynamics involves the generation of the phase fluctuation $\delta\theta(R)$, which along ${\bf e}_{\phi}$ direction is determined by 
\begin{equation}\label{PS1}
  (p_s^{\phi})^2=\!\!\sum_{q}q^2\Big[\frac{U_{q{\bf e}_{\phi}}U_{-q{\bf e}_{\phi}}}{4C\omega_N}+\frac{2n_B(\omega_N)\!+\!1}{2\omega_N}\Big(\frac{1\!+\!2DV_q}{D}\Big)\Big].
\end{equation}
Here, $p_s^{\phi}{\bf e}_{\phi}$ denotes the generated superconducting momentum ${\bf p}_s={\bm \nabla}_{\bf R}\delta\theta(R)/2$ by phase fluctuation along ${\bf e}_{\phi}$ direction; $U_{\bf q}$ represents the Fourier component of disorder-induced local electric potential; $n_B(x)$ stands for the Bose distribution; {\small $\omega_N=\sqrt{\omega_p^2+n_sq^2/(2Dm)}$} denotes the energy spectrum of the phase fluctuation, where $\omega_p=\sqrt{q^2V_{q}n_s/m}$ represents the plasma frequency and $n_s$ stands for the superfluid density, with $V_{q}$ being the Coulomb potential; $m$ and $D$ denote the effective mass and density of states of carriers, respectively; $1/C$ is the normalized factor in frequency-momentum space.

The microscopic electronic-fluid dynamics includes: the anomalous correlation $F_{\bf k}$, used to characterize the pairing and determine gap $\Delta_{\bf k}$ equation and superfluid density $n_s$; the microscopic momentum-relaxation rate $\Gamma_{\bf k}$ in superfluid. The expressions of these quantities are written as
\begin{eqnarray}\label{ac}
F_{\bf k}&=&\frac{f(E_{\bf k}^{+})-f(E_{\bf k}^-)}{2E_{\bf k}},\\
\label{GE1}
\Delta_{\bf k}&=&-{\sum_{\bf k'}}'g_{\bf kk'}{\Delta_{\bf k'}}F_{\bf k'},\\
n_s&=&\frac{k_F^2}{m}{\sum_{\bf k}}'\frac{\Delta^2_{\bf k}}{E_{\bf k}}\partial_{E_{\bf k}}{F_{\bf k}},\label{SD}\\
\Gamma_{\bf k}\!&=&\!-\!{\sum_{{\bf k'}\eta=\pm}}'|M_{\bf kk'}|^2\delta(E^{\eta}_{\bf k}-E^{-\eta}_{\bf k'}). \label{SCF}  
\end{eqnarray} 
Here, $E^{\pm}_{\bf k}$ denotes the quasi-electron and quasi-hole energies; $g_{\bf kk'}$ represents the pairing potential, in which we approximately taking $k=k'=k_F$ around the Fermi surface so that the gap $\Delta_{\bf k}$ only has angular dependence of the momentum; $\sum_{\bf k}'$ here and hereafter stands for the summation restricted in the spherical shell ($\xi_{\bf k}\le\omega_D$) with $\omega_D$ being the cutoff frequency, following the BCS theory\cite{BCS,G1}; $M_{\bf kk'}$ denotes the effective matrix element of the electron-impurity scattering. It is established that in the presence of the superconducting momentum ${\bf p}_s$, the quasiparticle energy is tilted as $E_{\bf k}^{\pm}=({\bf v}_{\bf k}\cdot{\bf p}_s)\pm{E_{\bf k}}$, where $E_{\bf k}=\sqrt{\xi_{\bf k}^2+\Delta_{\bf k}^2}$ is the original Bogoliubov quasiparticle energy and ${\bf v}_{\bf k}\cdot{\bf p}_s$ denotes the Doppler shift\cite{FF1,FF4,FF5,FF6,GIKE1}, with the group velocity ${\bf v}_{\bf k}=\partial_{\bf k}\xi_{\bf k}$. In the present work, we approximately take the parabolic spectrum, i.e., $\xi_{\bf k}={\bf k}^2/(2m)-\mu$ with $\mu$ being the chemical potential.

Particularly, it is noted that the generation of the phase fluctuation in Eq.~(\ref{PS1}) is coupled with electronic fluids through the superfluid density $n_s$ in the energy spectrum $\omega_N$, whereas the phase fluctuation is involved in the electronic-fluid dynamics in Eqs.~(\ref{ac})-(\ref{SCF}) by Doppler shift. The macroscopic phase-coherence dynamics and microscopic electronic-fluid dynamics are therefore mutually coupled. For phase fluctuation along a certain direction, by self-consistently solving Eqs.~(\ref{PS1}) and (\ref{GE1}) as well as (\ref{SD}), one can uniquely determine the superconducting gap $\Delta_{\bf k}$, superfluid density $n_s$ and superconducting phase fluctuation $p_s$, and then, the microscopic momentum-relaxation rate $\Gamma_{\bf k}$ and various physical quantities are obtained. Nevertheless, in realistic situation, there exist phase fluctuations along all directions. Hence, the experimentally observed quantity is a statistical average of phase fluctuations in all directions. The expected value of the physical quantity $X$ therefore reads
\begin{equation}\label{FAF}
  \langle{X}\rangle=\frac{1}{2\pi}{{\int}d\phi}X(\phi),
\end{equation}
with $X(\phi)$ being the solved $X$ for phase fluctuation along ${\bf e}_{\phi}$ direction. 

Equations~(\ref{PS1})-(\ref{SCF}) then provide an efficient and simplified way to understand the superconductivity properties in cuprate and disordered $s$-wave superconductors with significant phase fluctuation.

\section{Application to $d$-wave cuprate superconductors}

In this section, we apply the developed dual dynamics into the $d$-wave cuprate superconductors. Without losing generality, we choose $d_{x^2-y^2}$-wave order parameter for analysis, i.e., {\small $\Delta_{{\bf k}}=\Delta\cos(\zeta\theta_{\bf k})$} with $\zeta=2$, and the pairing potential $g_{\bf kk'}{\approx}g\cos[\zeta(\theta_{\bf k}-\theta_{\bf k'})]$ as a consequence of the translational and time-reversal symmetries\cite{SM1}. 

\subsection{Theoretical description of preformed Cooper pairs}
\label{sec-TD}

In this part, by simply performing an analytic analysis on the coupled dual dynamics of macroscopic phase coherence and microscopic electronic fluids, we present the physical pictures of the preformed Cooper pair model in pseudogap state (i.e., separation between $T_c$ and $T^{\rm os}$)  and the emerged normal fluid in superconducting one in $d$-wave cuprate superconductors. 

We start with the influence on the microscopic electronic fluids from the macroscopic phase fluctuation. Specifically, we first focus on the anomalous correlation. At low temperature, for an excited superconducting fluctuation ${\bf p}_s=p_s^{\phi}{\bf e}_{\phi}$, it is noted that considering the fact that $E_{\bf k}^+~{\ge}~E_{\bf k}^-$, the anomalous correlation $F_{\bf k}$ in Eq.~(\ref{ac}) vanishes in regions with $|{\bf v}_{\bf k}\cdot{\bf p}_s|>E_{\bf k}$, where the quasielectron energy $E_{\bf k}^+=({\bf v}_{\bf k}\cdot{\bf p}_s)+{E_{\bf k}}<0$ or quasihole energy $E_{\bf k}^-=({\bf v}_{\bf k}\cdot{\bf p}_s)-{E_{\bf k}}>0$, whereas $F_{\bf k}$ in regions with $|{\bf v}_{\bf k}\cdot{\bf p}_s|<E_{\bf k}$ is always finite. It has been established in the literature\cite{G1,FF1,GIKE1} that the nonzero anomalous correlation directly reflects the existence of the pairing as the characteristic quantity. Regions with nonzero and vanishing anomalous correlation are therefore referred to as the pairing and unpairing regions\cite{FF1,FF2,FF7,FF8,FF9,GIKE1}, respectively. Particularly, particles in the unpairing region no longer participate in the pairing and behave like the normal ones, leading to the emergence of the normal fluid\cite{GIKE1}. Interestingly, it is noted that due to the anisotropy of $d$-wave gap, the condition $|{\bf v}_{\bf k}\cdot{\bf p}_s|>E_{\bf k}$ of the unpairing region is always satisfied around the nodal points, irrelevant of the direction of the phase fluctuation. One therefore always finds a nonzero fraction of the normal fluid in cuprate superconductors even at low temperature, in consistency with the experimentally observed substantial fraction of the normal state far below $T_c$\cite{NSB1,NSB2}. This is very different from the $s$-wave superconductors, where the emergence of the normal fluid requires ${\bf p}_s>\Delta/v_F$ by the unpairing-region condition $|{\bf v}_{\bf k}\cdot{\bf p}_s|>E_{\bf k}$, leading to a threshold to realize normal fluid as revealed in our previous work\cite{GIKE1}.
  
Particles in the pairing regions with nonzero anomalous correlation contribute to the gap and superfluid density as superfluid. Moreover, according to the microscopic scattering of the momentum relaxation in superfluid [Eq.~(\ref{SCF})], one can further divide the pairing region into two parts: viscous one, in which $\Gamma_{\bf k}\ne0$; non-viscous one where $\Gamma_{\bf k}=0$. Specifically, in Eq.~(\ref{SCF}),  if ${\bf k}$ particle lies in the pairing region, one has $E_{\bf k}^+>0$ and $E_{\bf k}^-<0$. Then, once the energy conservation is satisfied to give rise to nonzero momentum-relaxation rate $\Gamma_{\bf k}$ of ${\bf k}$ particle, one finds $E_{\bf k'}^->0$ by $\delta(E_{\bf k}^+-E_{\bf k'}^-)$ or $E_{\bf k'}^+<0$ by $\delta(E_{\bf k}^--E_{\bf k'}^+)$, and hence, ${\bf k'}$ particle lies in the unpairing region. This scattering between particles in pairing and unpairing regions, behaves like the friction between superfluid and normal fluid, leading to the viscous superfluid with finite momentum-relaxation rate $\Gamma_{\bf k}$.  But if the energy conservation can not be satisfied for any ${\bf k}'$, the ${\bf k}$ particle is free from the momentum-relaxation scattering, and one therefore gets the non-viscous superfluid with zero momentum-relaxation rate.

\begin{widetext}
  \begin{center}
 \begin{figure}[htb]
   {\includegraphics[width=17.5cm]{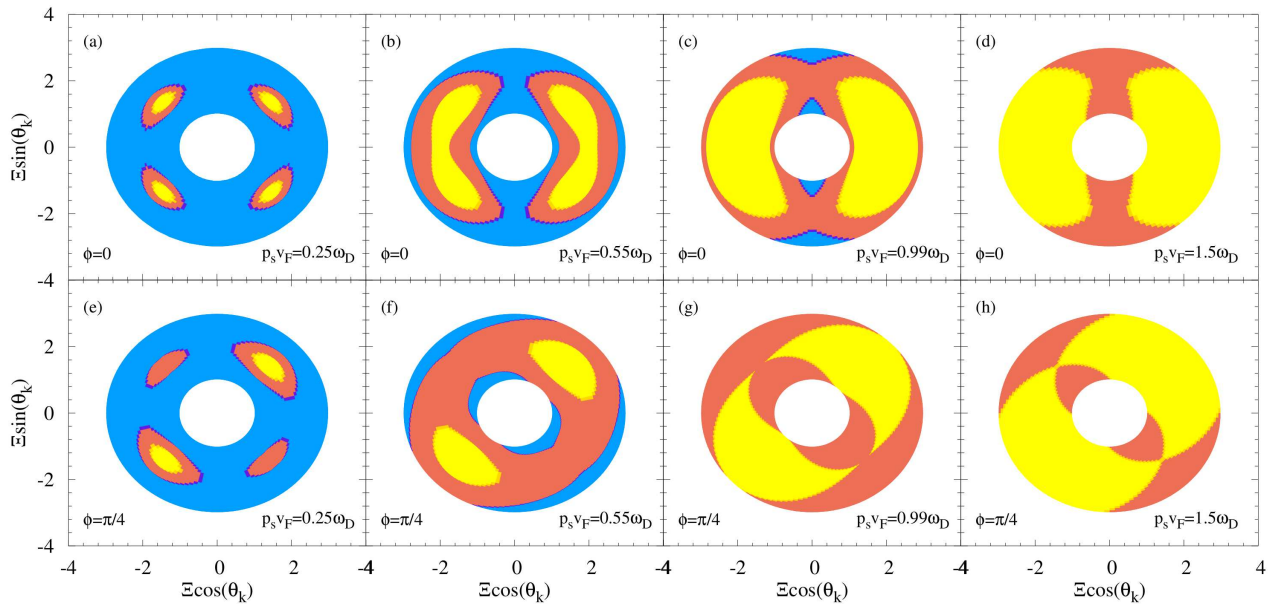}}
   \caption{(Color online)  Schematic showing the division in momentum space at different phase fluctuations ${\bf p}_s$. In the figure, $\Xi=\xi_{\bf k}+2\omega_D$ and hence the original spherical shell is characterized by $\omega_D\le\Xi\le3\omega_D$. The spherical shell is divided into three parts: unpairing region where anomalous correlation $F_{\bf k}=0$, non-viscous pairing region with $F_{\bf k}\ne0$ and momentum-relaxation rate $\Gamma_{\bf k}=0$, viscous pairing region with both $F_{\bf k}$ and $\Gamma_{\bf k}$ being finite, represented by yellow, blue and orange regions, respectively. (a)-(d): ${\bf p}_s$ along antinodal point ($\phi=0$); (e)-(h): ${\bf p}_s$ along nodal point ($\phi=\pi/4$). $\Delta=0.5\omega_D$ and $T=0$. }  
 \label{figyw1}
 \end{figure}
 \end{center}
\end{widetext}

Therefore, there exist three electronic fluids at small phase fluctuation in $d$-wave cuprate superconductors: normal fluid, viscous and non-viscous superfluids, similar to the three-fluid model proposed in our previous work\cite{GIKE1} in conventional superconductors which is caused by external electromagnetic field. Consequently, in analogy to Eq.~(\ref{SD}), we define the non-viscous and viscous superfluid densities as
\begin{equation}\label{nsd}
n_{\rm ns}=\frac{k_F^2}{m}{\sum_{{\bf k}\in{P_{\rm nv}}}}'\frac{\Delta^2_{\bf k}}{E_{\bf k}}\partial_{E_{\bf k}}F_{\bf k},  
\end{equation}
and
\begin{equation}\label{vsd}
n_{\rm vs}=\frac{k_F^2}{m}{\sum_{{\bf k}\in{P_{\rm v}}}}'\frac{\Delta^2_{\bf k}}{E_{\bf k}}\partial_{E_{\bf k}}F_{\bf k},  
\end{equation}
respectively. Whereas the normal-fluid density reads
\begin{equation}\label{nd}
n_{n}={\sum_{{\bf k}\in{U}}}'1.  
\end{equation}
Here, ${\bf k}\in{P_{\rm nv}}$, ${\bf k}\in{P_{\rm v}}$ and ${\bf k}\in{U}$ denote the summations restricted in the non-viscous ($F_{{\bf k}\in{P_{\rm nv}}}\ne0$ and $\Gamma_{{\bf k}\in{P_{\rm nv}}}=0$) and viscous ($F_{{\bf k}\in{P_{\rm v}}}\ne0$ and $\Gamma_{{\bf k}\in{P_{\rm v}}}\ne0$)  pairing regions as well as unpairing regions ($F_{{\bf k}\in{U}}=0$), respectively. Considering the short-circuit effect, if there exists the non-viscous pairing region in the spherical shell of the momentum space, i.e., non-viscous superfluid density, the system lies in the superconducting state, showing the zero-resistance (superconductivity) phenomenon.

It has been shown in Ref.~\onlinecite{GIKE1} that in conventional superconductors, the increase of Doppler shift leads to the increases of normal fluid and viscous superfluid but the shrinkage of the non-viscous superfluid. When the non-viscous superfluid vanishes at large enough Doppler shift, one can find an exotic state with only normal fluid and viscous superfluid left, showing the nonzero resistivity as well as the finite gap from the viscous superfluid. Following the same idea, in the spherical shell of the momentum space, a schematic illustration for the division of the unpairing, non-viscous and viscous pairing regions in $d$-wave superconductors is plotted in Fig.~(\ref{figyw1}). As seen from Figs.~\ref{figyw1}(a)-(d) [or Figs.~\ref{figyw1}(e)-(h)], the increase of ${p}_s$ first gradually enlarges the unpairing region (yellow regions) and hence viscous pairing region (orange regions) in the spherical shell, leading to the shrinkage of the non-viscous pairing region (blue regions). Particularly, once the non-viscous pairing region vanishes when the phase fluctuation exceeds a critical point, as shown in Fig.~\ref{figyw1}(d) [Fig.~\ref{figyw1}(g)], only viscous pairing and unpairing regions are left. Then, the system enters the pseudogap state with nonzero resistivity and finite gap due to the significant phase fluctuation. Correspondingly, in this circumstance, the viscous superfluid matches the description of the incoherent preformed Cooper pairs, as they both contribute to gap but experience the scattering. Whereas the existing normal fluid in our description implies the existence of normal particles in pseudogap state, which has been overlooked in previous preformed Cooper-pair model to describe the pseudogap state\cite{phase2,PCP1,PCP2,PCP3,PCP4,PCP5}. By further increasing ${p}_s$ in pseudogap state, the viscous pairing region starts to shrink, as shown in Figs.~\ref{figyw1}(g) and (h), until vanishes. The system with only unpairing region left eventually enters the normal state.  

In addition, in comparison between Figs.~\ref{figyw1}(a)-(d) and Figs.~\ref{figyw1}(e)-(h) for phase fluctuations along antinodal and nodal points, respectively, one finds that the influence of the phase fluctuation is anisotropic with respect to its direction. This anisotropy in fact arises from the anisotropy of the $d$-wave gap, similar to the observed anisotropic transport properties in the previous experiment\cite{Ro}. Specifically, at the same condition, in comparison to ${\bf p}_s$ along the antinodal point, ${\bf p}_s$ along the nodal point is easier to satisfy $|{\bf v}_{\bf k}\cdot{\bf p}_s|>E_{\bf k}$, and hence, causes the larger unpairing region, making it earlier to enter the pseudogap and then normal states with the gradual increase of $p_s$, as shown by comparison between Figs.~\ref{figyw1}(c) and~(g). Nevertheless, the judgment to enter the pseudogap and normal states in realistic situation is given by the vanishing non-viscous and viscous superfluid densities after the statistical average of the phase fluctuations in all directions, respectively.

We next discuss the influence on the macroscopic phase fluctuation from the microscopic superfluid. For the source terms on the right-hand side of Eq.~(\ref{PS1}), the first term arises from the disorder-induced local electric potential, which excites the inhomogeneous phase fluctuation through Josephson effect\cite{Josephson}. The second term comes from the zero-point energy and thermal excitation of the Bosonic phase fluctuation. Particularly, differing from the conventional bulk superconductors where the high-energy plasma frequency and hence $\omega_N\approx\omega_p$ causes the inactive phase fluctuation (Anderson-Higgs mechanism)\cite{AHM,AK,Am0,Ba0}, in consideration of the two-dimensional $V_{q}=2\pi{e^2}/({q\varepsilon})$, $\omega_N$ retains gapless in cuprate superconductors thanks to the layered structures\cite{cuprate4,cuprate5}, leading to an active phase fluctuation. 

For the convenience of analysis, by using the fact that {\small $\omega_N=\sqrt{\omega_p^2+n_sq^2/(2Dm)}=\sqrt{n_s/m}\sqrt{q^2V_{q}+q^2/(2D)}$}, we transform Eq.~(\ref{PS1}) into an equivalent form:
\begin{equation}\label{PSE}
(p_s^{\phi})^2\sqrt{n_s/m}={\cal{T}_{\phi}},  
\end{equation}
with ${\cal{T}}_{\phi}=\sum_{q}\big[\frac{q^2U_{q{\bf e}_{\phi}}U_{-q{\bf e}_{\phi}}/C+2q^2/D_q+4q^2n(\omega_N)/D_q}{4\sqrt{q^2V_{q}+q^2/(2D)}}\big]$. It is noted that the leading contribution of $n(\omega_N)$ lies in the long-wave regime, leading to the marginal role of $q^2n(\omega_N)$ in ${\cal{T}_{\phi}}$. In this circumstance, ${\cal{T}_{\phi}}$ contributed by disorder and zero-point-energy effects is independent of the phase fluctuation and electronic fluids as well as the temperature, and hence, acts as a structure factor in the phase-coherence dynamics in Eq.~(\ref{PSE}). Consequently, from Eq.~(\ref{PSE}), the factor $n_s/m$ (superfluid density over effective mass) plays a crucial role in determining the superconducting phase stiffness, i.e., an enhancement of ${n_s/m}$ suppresses the phase fluctuation, as the early experiments in cuprate superconductors realized\cite{phase1,phase2,phase3}.

In fact, the main thermal effect in the phase-coherence dynamics arises from the coupling to the electronic fluids. Specifically, with the increase of the temperature, the suppressed order parameter and hence the superfluid density [Eq.~(\ref{SD})] enlarge the phase fluctuation by weakening the phase stiffness. Once the temperature exceeds a critical point to generate the significant phase fluctuation, the system enters the pseudogap state with only viscous superfluid and normal fluid left. With further increase of temperature, the order parameter and hence superfluid density tend to vanish, and then, $p_s$ moves towards an infinitely large value, causing the unpairing region (i.e., normal fluid) left alone. Particularly, it is noted that the increased $p_s=\nabla_{\bf R}\delta\theta$ with temperature directly suggests a dephasing process of superconducting phase coherence from long range to short range.

\subsection{Numerical Simulation}
\label{sec-NS}

In this part, we perform a full numerical simulation to verify the theoretical analysis in the previous subsection. For the quantum effect of disorder on the macroscopic phase-coherence dynamics in Eq.~(\ref{PS1}), we apply the method of the Anderson disorder, which is introduced by generating random on-site potential at the structural sites, i.e., $U({\bf R}_i)=\gamma_iW$, where $\gamma_i$ denotes on-site random number with uniform probability in range $(-1,1)$ and $W$ represents the Anderson-disorder strength.
 Considering the crystal structure of cuprate superconductors, we take a finite square lattice system of large size $400\times400$ with the periodic boundary condition. 
 Then, in each random configuration, $U_{\bf q}=\sum_{{\bf R}_i}e^{i{\bf q}\cdot{\bf R}_i}U({\bf R}_i)$ in Eq.~(\ref{PS1}).
 
 In each random configuration, we self-consistently solve the phase fluctuation [Eq.~(\ref{PS1})] and gap equation [Eq.~(\ref{GE1})] as well as superfluid density [Eq.~(\ref{SD})] for phase fluctuation along a certain direction. Then, with the solved ${\bf p}_s$ and $\Delta$, by determining the unpairing region as well as non-viscous and viscous pairing regions according to the anomalous correlation [Eq.~({\ref{ac}})] and microscopic momentum-relaxation rate [Eq.~(\ref{SCF})], we can calculate the densities of the non-viscous [Eq.~(\ref{nsd})] and viscous [Eq.~(\ref{vsd})] superfluids and normal-fluid [Eq.~(\ref{nd})]. After that, by varying the direction of the phase fluctuation, we calculate the various quantities for phase fluctuation along each direction, and then, 
take the statistical average of the phase fluctuations in all directions [Eq.~(\ref{FAF})]. Finally, the gap $\langle\Delta\rangle$ as well as densities of the normal-fluid $\langle{n_{n}}\rangle$, viscous $\langle{n_{\rm vs}}\rangle$ and non-viscous $\langle{n_{\rm ns}}\rangle$ superfluids are averaged over $400$ random configurations for convergence. Moreover, in the simulation, the gap at zero temperature in the absence of the phase fluctuation is taken to be $\Delta_0=22~$meV, which is close to the maximum value of the observed gap in YBa$_2$Cu$_3$O$_{7-x}$\cite{Ypa}. $C=2\omega_D/\xi^{2}_c$ with $\xi_c=\hbar{v_F}/\Delta_0$ being the coherence length\cite{G1}. The Fermi energy $E_F=220~$meV and $m_{\rm eff}=1.9m_0$\cite{Ypa1,Ypa2}, with $m_0$ representing the free-electron mass. The specific cutoff frequency $\omega_D$ requires the microscopic pairing mechanism, which still remains an open question in the literature. We chose $\omega_D=\Delta_0$, and then, due to the phase fluctuation, one has $\Delta\ll\omega_D$ in the simulation. With determined $\Delta_0$ and $\omega_D$, the pairing potential $g$ is determined by gap equation [Eq.~(\ref{GE1})] at zero temperature in the absence of the phase fluctuation. Furthermore, our model focuses on the long-wave (i.e., low-frequency) regime (refer to Appendix~\ref{app2}), and hence, we introduce a cutoff $q_c$ in the summation of $q$ in Eq.~(\ref{PS1}) to approximately guarantee that the energy spectrum of the phase fluctuation does not enter the Bogoliubov quasiparticle continuum along antinodal points, i.e., $\omega_N(q_c)<2\Delta_0$.

\begin{widetext}
  \begin{center}
 \begin{figure}[htb]
   {\includegraphics[width=17.5cm]{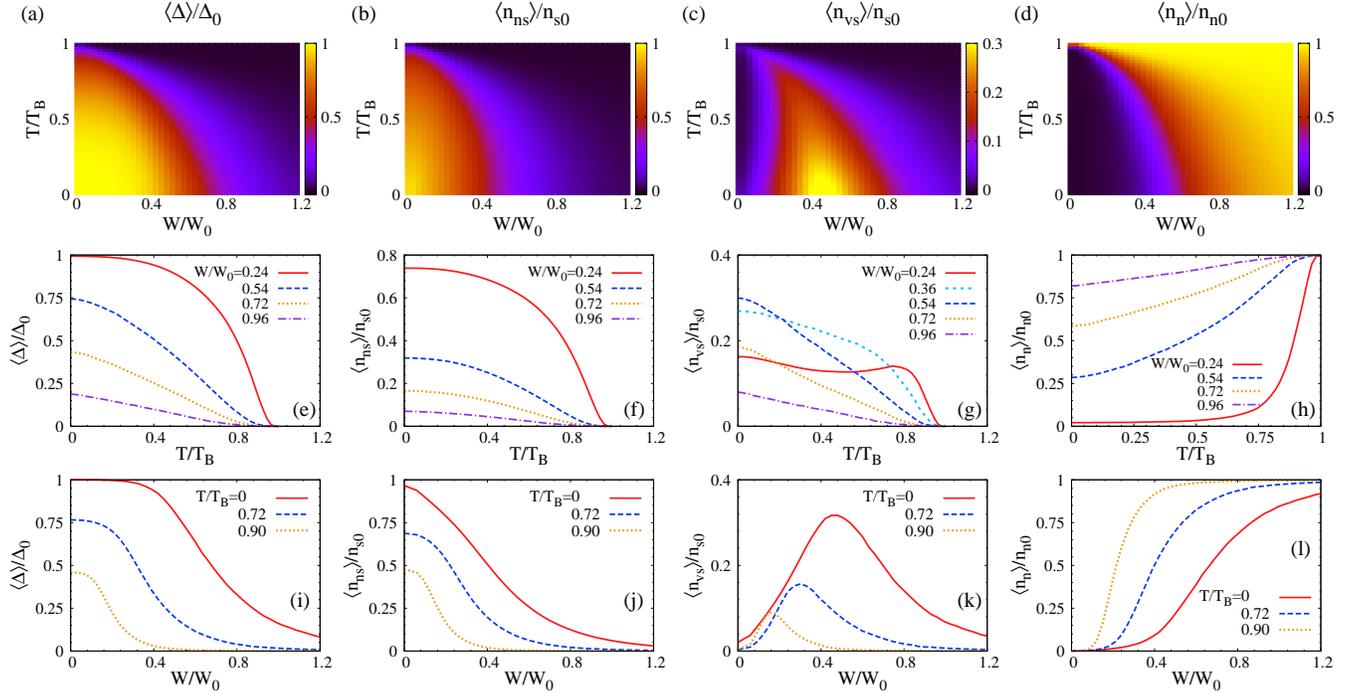}}
   \caption{(Color online) The temperature and Anderson-disorder strength dependence  of the expected value of the gap $\langle\Delta\rangle$ as well as densities of the non-viscous superfluid $\langle{n_{\rm ns}}\rangle$ , viscous superfluid $\langle{n_{\rm vs}}\rangle$ and normal fluid $\langle{n_{n}}\rangle$ after the statistical average of the phase fluctuations in all directions from the full and self-consistent numerical simulation. (a)-(d): phase diagram; (e)-(h): temperature dependence at different Anderson-disorder strengths; (i)-(l): Anderson-disorder strength dependence at different temperatures. In the figure, $T_{B}$ denotes the critical temperature where gap vanishes in the absence of the phase fluctuation, i.e., BCS critical temperature of two-dimensional $d$-wave superconductors; $n_{s0}$ represents the superfluid density at zero temperature in the absence of the phase fluctuation, and $n_{n0}$ denotes the normal-fluid density above $T^{\rm os}$; $W_0=0.5\Delta_0$. }  
 \label{figyw2}
 \end{figure}
 \end{center}
\end{widetext}
 
\subsubsection{Phase Diagram}
\label{PDsec}

The temperature and Anderson-disorder strength dependence of the expected values of the gap as well as densities of the normal-fluid, viscous and non-viscous superfluids after the statistical average of the phase fluctuations in all directions are plotted in Fig.~\ref{figyw2}.

We first discuss the gap. As seen from the phase diagram in Fig.~\ref{figyw2}(a), $\langle\Delta\rangle$ decreases with the increase of the Anderson-disorder strength $W$ or temperature $T$, exhibiting a half-dome behavior. Specifically, with the increase of temperature, as shown in Fig.~\ref{figyw2}(e), the gap shows the BCS-like behavior at small $W$ (red solid curve) as it should be, but tends to become linear decrease when $W$ is enhanced (ochre dotted and purple chain curves). Particularly, the change into the linear decrease starts at high-$T$ regime, and moves towards low-$T$ one with the enhancement of $W$.
This temperature dependence from our numerical simulation qualitatively agrees with the previous experimental observation\cite{STM4} where the temperature dependence of the gap shows a small platform in low-$T$ regime but linearly decreases in the remaining temperature regime. The deviation of the temperature dependence from the BCS-like behavior here arises from the phase fluctuation. By increasing temperature, the suppressed gap and hence superfluid density (especially in high-$T$ regime) enhances the phase fluctuation and hence Doppler shift, and then, the shrinkage of the pairing region as a result feeds back to suppress gap, speeding up the gap falling in comparison to BCS-like behavior. This effect by phase fluctuation can be more directly seen from the $W$ dependence in Fig.~\ref{figyw2}(i). By increasing $W$ from zero, $\langle\Delta\rangle$ first changes marginally, and then, exhibits an exponential decay at relatively large $W$ where phase fluctuation becomes important. 

By comparing Figs.~\ref{figyw2}(a) and~(b), one finds that the normalized non-viscous superfluid density $\langle{n_{\rm ns}}\rangle/n_{s0}$ also exhibits a half-dome behavior in phase diagram, but is smaller than that of the normalized gap $\langle\Delta\rangle/\Delta_0$. By further comparing Figs.~\ref{figyw2}(e) and~(f) as well as Figs.~\ref{figyw2}(i) and~(j), with the increase of $T$ or $W$, $\langle{n_{\rm ns}}\rangle/{n_{s0}}$ exhibits a faster decrease than $\langle\Delta\rangle/\Delta_0$. This is because that in comparison to $\langle{n_{\rm ns}}\rangle$ contributed by non-viscous pairing region alone, both viscous and non-viscous pairing regions contribute to the gap. Therefore, with the enhancement of $p_s$ by increasing $T$ or $W$, the faster shrinkage of the non-viscous pairing region than the entire pairing region (as shown in Fig.~\ref{figyw1}) leads to the faster decrease of $\langle{n_{\rm ns}}\rangle/n_{s0}$ than $\langle\Delta\rangle/\Delta_0$. We point out that the non-viscous superfluid density can be directly measured by detecting the $1/\omega$-like divergent
behavior in the imaginary part of the optical conductivity [i.e., $\sigma_2(\omega)=\frac{e^2{\langle}n_{\rm ns}{\rangle}}{m\omega}$] at low-frequency regime\cite{Infrad2,Infrad3,MS1,MS3,THZ3,THZ4,NSB2}. 

The normal-fluid density in Figs.~\ref{figyw2}(d), (h) and (l) shows compensatory behavior in comparison to the non-viscous superfluid density in corresponding Figs.~\ref{figyw2}(b), (f) and (j), as it should be according to the analytic analysis in Sec.~\ref{sec-TD}. Particularly, the compensatory behavior between the superfluid and normal-fluid densities in the temperature dependence has been observed in the previous experiment\cite{NSB2}. Moreover, it is noted in Fig.~\ref{figyw2}(l) that at relatively large $W$ (i.e., phase fluctuation), the nonzero normal-fluid density at zero temperature (red solid curve) agrees with our analysis in Sec.~\ref{MDDsec}, in consistency with the previous experimental observation\cite{NSB1,NSB2}.

The viscous superfluid density $\langle{n_{\rm vs}}\rangle$ exhibits very differently, showing a flame-like behavior in phase diagram in Fig.~\ref{figyw2}(c). From Eq.~(\ref{vsd}), there exist two opposite effects on $\langle{n_{\rm vs}}\rangle$ by increasing phase fluctuation: (i) at small phase fluctuation, through the friction, the increase of the unpairing region enlarges the viscous pairing one and hence $\langle{n_{\rm vs}}\rangle$; (ii) the suppressed gap directly reduces $\langle{n_{\rm vs}}\rangle$. As shown in Fig.~\ref{figyw2}(k), with the increase of $W$, effects (i) and (ii) dominate at small and large $W$, leading to 
the increase and decrease of ${\langle}n_{\rm vs}\rangle$, respectively. A peak is therefore observed, and the peak position moves towards small $W$ with the increase of temperature as effect (ii) is enhanced. As for the temperature dependence in Fig.~\ref{figyw2}(g), at small $W$ (red solid curve), with the increase of temperature, a platform due to the competition between effects (i) and (ii) is found in low-$T$ regime, whereas effect (ii) dominates around the critical temperature where the gap changes dramatically, leading to the decrease of $\langle{n_{\rm vs}}\rangle$. At large $W>0.5$, effect (ii) dominates, causing the decrease of $\langle{n_{\rm vs}}\rangle$.

\begin{figure}[htb]
   {\includegraphics[width=8.5cm]{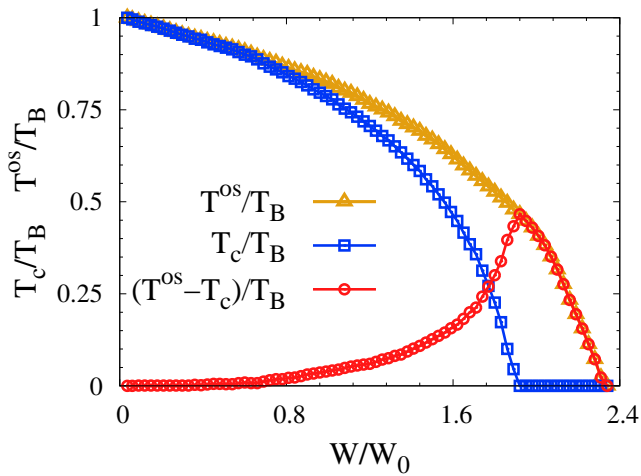}}
   \caption{(Color online) Critical temperatures $T_c$ and $T^{\rm os}$ as well as the difference between $T^{\rm os}$ and $T_c$ versus Anderson-disorder strength. In the figure, $T_{B}$ denotes the critical temperature where gap vanishes in the absence of the phase fluctuation, i.e., BCS critical temperature of two-dimensional $d$-wave superconductors. $W_0=0.5\Delta_0$.}  
 \label{figyw3}
 \end{figure}

Now we propose a scheme to detect the specific viscous superfluid density $\langle{n_{\rm vs}}\rangle$. Following our previous work in conventional superconductors\cite{GIKE1}, based on three-fluid model, in the linear optical response, the optical conductivity in low-frequency regime reads
\begin{equation}\label{TFOB}
  \sigma(\omega)=\frac{e^2\langle{n_{\rm ns}}\rangle}{im\omega}+\frac{e^2\langle{n_{\rm vs}}\rangle}{m(i\omega+\gamma_{\rm vs})}+\frac{e^2\langle{n_{n}}\rangle}{m(i\omega+\gamma_{n})},
\end{equation}
with $\gamma_{\rm vs}$ and $\gamma_{n}$ being the relaxation rates of viscous superfluid and normal fluid, respectively. Here, the normal fluid (the third term) exhibits the well-known Drude-model behavior; the non-viscous superfluid (the first term) is free from the resistance; due to the friction between superfluid and normal fluid, the viscous superfluid (the second term) also shows the Drude-model behavior. Therefore, as mentioned above, $\langle{n_{\rm ns}}\rangle$ can be directly measured by detecting the $1/\omega$-like divergent behavior in the imaginary part of the optical conductivity at low-frequency regime\cite{Infrad2,Infrad3,MS1,MS3,THZ3,THZ4,NSB2}.  Whereas from Eq.~(\ref{TFOB}), one has 
\begin{equation}
\int^{\infty}_{0+}d\omega\sigma_1(\omega)=\frac{2e^2}{{\pi}m}(\langle{n_{\rm vs}}\rangle+\langle{n_{n}}\rangle),  
\end{equation}
with $\sigma_1(\omega)$ being the real part of the optical conductivity. Therefore, by detecting the area under $\sigma_1(\omega)$ curve in frequency dependence\cite{NSB2}, $\langle{n_{\rm vs}}\rangle+\langle{n_{n}}\rangle$ is obtained. 

Furthermore, in the magnetic response, based on three-fluid model, the excited current is written as\cite{GIKE1}
\begin{equation}\label{mc}
{\bf j}=-\frac{e^2{\bf A}\langle{n_{\rm ns}}\rangle}{m}-\frac{e^2{\bf A}\langle{n_{\rm vs}}\rangle}{m}\Big(1-\frac{\xi_c}{l}\Big)+\frac{e^2{\bf A}\langle{n_{\rm n}}\rangle}{m}\frac{\xi_c}{l},   
\end{equation}
with $\xi_c$ and $l$ being the coherence length and mean-free path, respectively. Here, the first and second terms come from the excited supercurrent in non-viscous and viscous superfluids by the Meissner effect\cite{G1}, respectively. It is noted that the viscous superfluid experiences the resistance due to the friction between superfluid and normal fluid. The magnetic flux can not drive the normal-fluid current directly, but through the friction drag with the superfluid current, a normal-fluid current (the third term, proportional to $1/l$) is excited. From Eq.~(\ref{mc}), the magnetic penetration depth $\lambda$ is determined by 
\begin{equation}
\frac{1}{\lambda^2}=\frac{e^2(\langle{n_{\rm ns}}\rangle+\langle{n_{\rm vs}}\rangle)}{m}-\frac{e^2(\langle{n_{\rm vs}}\rangle+\langle{n_{\rm n}}\rangle)}{m}\frac{\xi_c}{l}.   
\end{equation}
Consequently, from above equation, with the established $\xi_c$ and $l$ as well as the obtained densities $\langle{n_{\rm vs}}\rangle+\langle{n_{n}}\rangle$ and $\langle{n_{\rm ns}}\rangle$ from the optical detection as mentioned above, one can determine the viscous superfluid density $\langle{n_{\rm vs}}\rangle$ and also the normal-fluid density $\langle{n_n}\rangle$ by measuring the penetration depth $\lambda$.

\subsubsection{Separation between $T_c$ and $T^{\rm os}$}

We next discuss the separation between $T_c$ and $T^{\rm os}$. In the full numerical simulation, $T_c$ and $T^{\rm os}$ are chosen at the critical temperatures where the normalized non-viscous superfluid density $\langle{n_{\rm ns}}\rangle/{n_{s0}}$ and normalized gap $\langle\Delta\rangle/\Delta_0$ vanish, respectively, and are plotted in Fig.~\ref{figyw3}.  As mentioned above, with the increase of temperature, the decrease of $\langle{n_{\rm ns}}\rangle/{n_{s0}}$ is faster than $\langle\Delta\rangle/\Delta_0$, due to the additional contribution in gap from the viscous pairing region. $T_c$ is therefore smaller than $T^{\rm os}$, as shown in Fig.~\ref{figyw3}. Whereas when $T^{\rm os}>T>T_c$, since the non-viscous superfluid vanishes, the system with only normal fluid and viscous superfluid left enters the pseudogap state, showing both nonzero resistivity and finite gap.

To enlarge the separation between $T_c$ and $T^{\rm os}$ (i.e., to lower $T_c$), one needs to enhance $W$ in order to generate significant phase fluctuation at low temperature, so that the system enters the pseudogap state at lower temperature. As seen from Fig.~\ref{figyw3}, the separation between $T_c$ and $T^{\rm os}$  at small $W$ is marginal. By increasing $W$, both $T^{\rm os}$ (ochre solid curve with triangles) and $T_c$ (blue solid curve with squares) decrease due to the suppressed gap [Fig.~\ref{figyw2}(i)] and non-viscous superfluid density [Fig.~\ref{figyw2}(j)], respectively. Whereas the separation between $T_c$ and $T^{\rm os}$ (red solid curve with circles) is enlarged until $T_c=0$, thanks to the enhanced phase fluctuation at low temperature. Particularly, with the large enough phase fluctuation at $1.9<W/W_0<2.3$, one can find the emergence of the pseudogap state even at zero temperature.

In our simulation with the superconducting-material parameters of YBa$_2$Cu$_3$O$_{7-x}$ ($\Delta_0=22~$meV and hence $T_B=110~$K)\cite{Ypa}, $T^{\rm os}$ tracks $T_c$ closely in Fig.~\ref{figyw3} ($35$~K above $T_c$ at most when $T_c>16~$K and $50~$K above $T_c$ at most when $T_c=0$), and lies well below the experimental pseudogap temperature $T^*$ in underdoped regime\cite{NF1}, in consistency with the current experimental findings and understanding\cite{TM2,TM3,TM4,SH,PC1,NF1,AF1,AF2,AF3,AF4,PPS1,PPS2,Infrad1,Infrad2,Infrad3,MS1,MS2,MS3,THZ1,THZ2,THZ3,THZ4,DHM3} mentioned in the introduction. This close track in fact arises from the fast shrinkage of the remaining viscous pairing region in the pseudogap state with the increase of temperature and hence phase fluctuation, as shown in Figs.~\ref{figyw1}(g) and (h), leading to a strong suppression on the gap. In other words, once in the pseudogap state, because of the significant phase fluctuation and hence remarkable Doppler shift, the gap can not survive far above $T_c$ in our simplified model.   

\begin{figure}[htb]
   {\includegraphics[width=8.0cm]{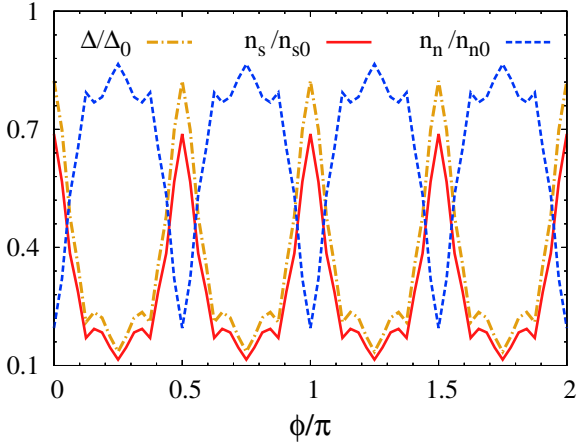}}
   \caption{(Color online) Dependence of the gap and densities of the normal fluid and superfluid on the phase-fluctuation direction after the average over $400$ random configurations. $W/W_0=0.72$ and $T/T_B=0.18$. In the figure, $n_{s0}$ represents the superfluid density at zero temperature in the absence of the phase fluctuation, and $n_{n0}$ denotes the normal-fluid density above $T^{\rm os}$.  $T_{B}$ denotes the critical temperature where gap vanishes in the absence of the phase fluctuation, i.e., BCS critical temperature of the two-dimensional $d$-wave superconductors.  $W_0=0.5\Delta_0$. }  
 \label{figyw4}
\end{figure}

\subsubsection{Anisotropic influence of phase fluctuation}

It is noted that in our numerical calculation, after the statistical average of the phase fluctuations in all directions, there is no current excitation, i.e., $\langle{\bf j}\rangle=0$, since $\langle{\bf p}_s\rangle=0$ as a consequence of the inversion symmetry in our system. Whereas the nonzero gap $\langle{\Delta}\rangle$ and densities of the normal-fluid $\langle{n_{n}}\rangle$, viscous $\langle{n_{\rm vs}}\rangle$ and non-viscous $\langle{n_{\rm ns}}\rangle$ superfluids are due to the fact that the anomalous correlation [Eq.~(\ref{ac})] and hence the gap equation [Eq.~(\ref{GE1})] and superfluid density [Eq.~(\ref{SD})], as well as the microscopic momentum-relaxation rate $\Gamma_{\bf k}$ [Eq.~(\ref{SCF})] are even functions with respect to ${\bf p}_s$, whereas $\langle{p_s^2}\rangle\ne0$ by Eq.~(\ref{PS1}).

In Fig.~\ref{figyw4}, after the average over $400$ random configurations, we plot the dependence of the gap and densities of the normal fluid and superfluid on the phase-fluctuation direction. As seen from the figure, all three quantities exhibit $C_4$-symmetric dependence on the phase-fluctuation direction. The superfluid density shows similar dependence to the gap, but exhibits compensatory behavior in comparison with the normal-fluid density, similar to the results in Sec.~\ref{PDsec}. The $C_4$-symmetric dependence with respect to the phase-fluctuation direction arises from the $d$-wave gap, since the phase fluctuation along the nodal point is easier to generate the unpairing region (normal fluid) and reduce the pairing region (i.e., suppress the gap and hence superfluid density) than that along the antinodal point, as analyzed in Sec.~\ref{MDDsec}. 

For experimental probes that are related to the intrinsic characters of system, the superconducting momentum ${\bf p}_s$ only arises from the phase fluctuation, and then, the $C_4$-symmetric anisotropies in Fig.~\ref{figyw4} does not manifest directly, since there exist phase fluctuations along all directions and the observed quantity is a statistical average of phase fluctuations in all directions. However, one can apply the external stationary magnetic vector potential ${\bf A}$ or inject the background supercurrent ${\bf I}$ to generate the extrinsic superconducting momentum ${\bf p}_s^{\rm ext}=-e{\bf A}$ or ${\bf p}_s^{\rm ext}={\bf I}m/(en_s)$. Consequently, the total superconducting momentum, including the intrinsic part from phase fluctuation and extrinsic one, is enhanced along the direction of ${\bf p}_s^{\rm ext}$. In this circumstance, by varying the direction of the vector potential or injected supercurrent, all $C_4$-symmetric anisotropies in Fig.~\ref{figyw4}, i.e, the anisotropic influence of ${\bf p}_s$ mentioned in Sec.~\ref{MDDsec} can be observed.

\section{Derivation of dual dynamics}
\label{sec-F}

In this section, we present the rigorous derivation of the coupled dual dynamics from the GIKE approach. The recovery from the path-integral approach is also addressed to confirm our derivation. 

\subsection{Hamiltonian}

We begin with a generalized Bogoliubov-de Gennes (BdG) Hamiltonian\cite{G1,ER}:
\begin{equation}
  H_0=\int{d{\bf r}}{d{\bf r'}}\psi^{\dagger}(x)[\xi_{\hat {\bf p}}\tau_3\delta(x-x')+{\hat \Delta}(x,x')]\psi(x').\label{Ham0}
\end{equation}
Here, $\psi(x)=[\psi_{\uparrow}(x),\psi^{\dagger}_{\downarrow}(x)]$ represents the field operator in Nambu space; the momentum operator ${\hat {\bf p}}=-i\hbar{\bm \nabla}$; ${\hat \Delta}(x,x')=\Delta(x,x')\tau_++\Delta^*(x,x')\tau_-$; $\tau_i$ are the Pauli matrices in Nambu space.

It is established that the phase fluctuation in Eq.~(\ref{Ham0}) can be effectively removed by a unitary transformation:
\begin{equation}\label{utrp}
\psi(x){\rightarrow}e^{i\tau_3\delta\theta(R)/2}\psi(x),
\end{equation}
and then, one has\cite{gi0,AK}
\begin{eqnarray}
  H_0&=&\int{d{\bf r}}{d{\bf r'}}\psi^{\dagger}(x)\Big\{(\xi_{{\hat{\bf p}}+{\bf p}_s\tau_3}+\partial_t\delta\theta/2)\tau_3\delta(x-x')\nonumber\\
 &&   \mbox{}+\sum_{\bf k}e^{i{\bf k}\cdot({\bf r-r'})}[\Delta_{{\bf k}}+\delta\Delta_{\bf k}(R)]\tau_1\Big\}\psi(x').\label{Ham1}
\end{eqnarray}
with the superconducting momentum ${\bf p}_s={\bm \nabla}_{\bf R}\delta\theta(R)/2$. Then, it is clearly seen that as a consequence of the spontaneous breaking of the $U(1)$ symmetry by order parameter\cite{gi0,AK}, the superconducting phase fluctuation provides an effective electromagnetic potential $eA^{\rm eff}_{\mu}=(\partial_t\delta\theta/2,-{\bf p}_s)$, in consistency with the gauge structure in superconductors first revealed by Nambu\cite{gi0}:
\begin{eqnarray}
eA_{\mu}&\rightarrow&eA_{\mu}-\partial_{\mu}\chi(R), \label{gaugestructure1}\\
\label{gaugestructure2}
\delta\theta(R)&\rightarrow&\delta\theta(R)+2\chi(R).
\end{eqnarray}
Here, the standard electromagnetic potential $eA_{\mu}=(e\phi,e{\bf
  A})$ and $\partial_{\mu}=(\partial_t,-{\bm \nabla}_{\bf R})$. 

The electron-electron Coulomb interaction $H_{\rm ee}$ and electron-impurity interaction $H_{\rm ei}$ are written as\cite{G1}
\begin{eqnarray}
  &H_{\rm ee}=\frac{1}{2}{\int}d{\bf r}d{\bf r'}V(x-x')[\psi^{\dagger}({x})\tau_3\psi({x})][\psi^{\dagger}({x}')\tau_3\psi({x}')],&\nonumber\\
  \\
&H_{\rm ei}=\int{d{\bf r}}\psi^{\dagger}(x)U(x)\tau_3\psi(x).&  
\end{eqnarray}
Here, $V(x-x')$ and $U(x)$ denote the Coulomb and impurity potentials in space-time coordinate. We have kept the time dependence of $U(x)$ for general situation.

\subsection{GIKE approach}

To derive the macroscopic phase-coherence dynamics and microscopic electronic superfluid dynamics, we first use the GIKE approach\cite{GIKE1,GIKE2,GIKE3,GIKE4}. In this microscopic approach, the response of the system is described by the density of matrix $\rho_{\bf k}$ in Nambu space. Considering the fluctuations, the density of matrix reads
\begin{equation}
\rho_{\bf k}=\rho_{\bf k}^{(0)}+\delta\rho_{\bf k}(R).
\end{equation}
Here, $\delta\rho_{\bf k}$ stands for the non-equilibrium response from the equilibrium part $\rho_{\bf k}^{(0)}$, which is written as\cite{GIKE1}
\begin{equation}\label{r0}
\rho_{\bf k}^{(0)}=F_{\bf k}(E_{\bf k}\tau_0+{\xi_{\bf k}}\tau_3+{\Delta_{\bf k}}\tau_1).  
\end{equation}

To determine the density of matrix, one needs to solve $\delta\rho_{\bf k}$ from the GIKE\cite{GIKE1,GIKE2,GIKE3,GIKE4}:
\begin{eqnarray}
&&\partial_t\rho_{\bf
  k}\!+\!i\left[\left(\xi_k\!+\!\mu_{\rm
      eff}\!+\!\frac{p^2_s}{2m}\right)\tau_3\!+\!\Delta_{\bf k}\tau_1,\rho_{\bf 
k}\right]\!-\!\Big[\frac{i\nabla^2_{\bf R}}{8m}\tau_3,\rho_{\bf 
      k}\Big]\nonumber\\
&&\mbox{}\!+\!\frac{1}{2}\left\{e{\bf E}\tau_3\!-\!({\bm \nabla}_{\bf R}\!+\!2i{\bf p}_s\tau_3)\Delta_{\bf k}\tau_1,\partial_{\bf k}\rho_{\bf
k}\right\}\!\nonumber\\
&&\mbox{}\!-\!\frac{i}{8}\left[({\bm \nabla}_{\bf R}\!+\!2i{\bf p}_s\tau_3)({\bm
    \nabla}_{\bf R}\!+\!2i{\bf p}_s\tau_3)\Delta_{\bf k}\tau_1,\partial_{\bf k}\partial_{\bf k}\rho_{\bf
k}\right]\nonumber\\
&&\mbox{}\!+\!\Big\{\frac{\bf k}{2m}\tau_3,{\bm \nabla}_{\bf R}\rho_{\bf
    k}\Big\}\!+\!\Big[\frac{{\nabla_{\bf R}}\circ{\bf p}_s}{4m}\tau_3,\tau_3\rho_{\bf
    k}\Big]\!=\!\partial_t\rho_{\bf k}\Big|_{\rm
scat},
\label{GE}
\end{eqnarray}
where we have effectively removed the phase fluctuation from the order parameter through the unitary transformation in Eq.~(\ref{utrp}). Here, $[A,B]=AB-BA$ and $\{A,B\}=AB+BA$ represent the commutator and anti-commutator, respectively; ${\nabla_{\bf R}}\circ{\bf p}_s=2{\bf p}_s\cdot{\bm \nabla}_{\bf R}+{\bm \nabla}_{\bf R}\!\cdot\!{\bf p}_s$; the effective field $\mu_{\rm eff}(R)={\partial_t\delta\theta(R)}/{2}+\mu_{H}(R)+U(R)$ with the Hartree field written as
\begin{equation}\label{HF}
  \mu_H(R)=\sum_{R'}V(R-R')n({R'}).
\end{equation}
The induced electric field $e{\bf E}=-{\bm \nabla}_{\bf R}[\mu_H(R)+U(R)]$.   

The density $n$ and current ${\bf j}$ are given by\cite{GIKE1,GIKE2,GIKE3,GIKE4}
\begin{eqnarray}
 &n=\sum_{\bf k}(1+2\rho_{{\bf k}3}),& \label{density} \\ 
&{\bf
  j}=\sum_{\bf k}\big(\frac{e{\bf k}}{m}\rho_{{\bf k}0}\big),&\label{current}
\end{eqnarray}
respectively. The equation of the order parameter reads
\begin{equation}
  \left(\begin{array}{cc}
    & \Delta_{\bf k}\!+\!\delta\Delta_{\bf k}\\
    \Delta_{\bf k}\!+\!\delta\Delta_{\bf k} &
  \end{array}\right)=-{\sum_{\bf k'}}'g_{\bf kk'}\left(\begin{array}{cc}
    & \rho_{{\bf k}'+}\\
    \rho_{{\bf k}'-} &
  \end{array}\right),  
\end{equation}
whose components are given by
\begin{eqnarray}
{\sum_{\bf k'}}'g_{\bf kk'}\rho_{{\bf k'}1}&=&-\Delta_{\bf k}\label{gap},\\
{\sum_{\bf k'}}'g_{\bf kk'}\rho_{{\bf k'}2}&=&0\label{phase}.
\end{eqnarray}
Here, $\rho_{{\bf k}i}$ stands for the $\tau_i$ component of $\rho_{{\bf k}}=\sum_{i=0}^4\rho_{{\bf k}i}\tau_i$; It is noted that Eq.~(\ref{gap}) gives the gap equation, whereas Eq.~(\ref{phase}) determines the phase fluctuation as revealed in our previous work~\cite{GIKE2}.

The impurity scattering $\partial_t\rho_{\bf k}\big|_{\rm scat}$ is derived based on the generalized Kadanoff-Baym ansatz with the random-phase and Markovian approximations\cite{GQ2,GQ3}. The specific impurity-scattering term reads\cite{GIKE1} 
\begin{eqnarray}\label{scat1}
\partial_t\rho_{\bf k}|_{\rm scat}&\!=\!&\!-n_i\pi\sum_{{\bf k'}\eta=\pm}|U^s_{\bf
  k\!-\!k'}|^2\delta(E_{k'}\!-\!E_{k})(\tau_3\Gamma^{\eta}_{\bf k'}\tau_3\Gamma^{\eta}_{\bf k}\rho_{\bf k}\nonumber\\
&&\mbox{}\!-\!\tau_3\rho_{\bf
  k'}\Gamma^{\eta}_{\bf k'}\tau_3\Gamma^{\eta}_{\bf k}\!+\!h.c.).\label{FDSC}
\end{eqnarray}
Here, $n_i$ and $U^s_{\bf kk'}$ stand for the impurity density and matrix element of the electron-impurity scattering, respectively; the projection operators {\small $\Gamma^{\pm}_{\bf 
    k}=\mathscr{U}_{\bf k}^{\dagger}(1\pm\tau_3)\mathscr{U}_{\bf k}/{2}$} with $\mathscr{U}_{\bf k}=u_{\bf k}\tau_0-iv_{\bf k}\tau_2$  being the unitary transformation matrix from the particle space to the quasiparticle one. $u_{\bf k}=\sqrt{1/2+\xi_{\bf k}/(2E_{\bf k})}$ and
$v_{\bf k}=\sqrt{1/2-\xi_{\bf k}/(2E_{\bf k})}$. 

To solve GIKE, by expanding {\small $\delta\rho_{\bf k}=\delta\rho_{\bf k}^{(1)}+\delta\rho_{\bf k}^{(2)}$} with {\small $\delta\rho_{\bf k}^{(1)}$} and {\small $\delta\rho_{\bf k}^{(2)}$} standing for the linear and second-order terms of the nonequilibrium response, the GIKE becomes a chain of equations,
 whose first order only involves {\small $\delta\rho^{(1)}_{\bf k}$} and {\small
  $\rho^{(0)}_{\bf k}$} and second order involves {\small
  $\delta\rho^{(2)}_{\bf k}$}, {\small $\delta\rho^{(1)}_{\bf k}$} and {\small
  $\rho^{(0)}_{\bf k}$}. Then, one can solve $\delta\rho_{\bf k}^{(1)}$ and $\delta\rho_{\bf k}^{(2)}$ in sequence, whose specific lengthy expressions are presented in Appendix~\ref{app1}.

\subsubsection{Gap equation and superfluid density}
\label{sec-GESD}

We first derive the gap equation and supercurrent. Substituting the solved $\rho_{{\bf k}1}=\rho_{{\bf k}1}^{(0)}+\delta\rho_{{\bf k}1}$ into Eq.~(\ref{gap}), with $\rho_{{\bf k}1}^{(0)}=\Delta_{\bf k}F_{\bf k}$ from Eq.~(\ref{r0}), one can directly obtain the gap equation in Eq.~(\ref{GE1}), and find a vanishing amplitude fluctuation $\delta\Delta_{\bf k}(R)=0$ as $\delta\rho_{{\bf k}1}$ makes no contribution (refer to Appendix~\ref{app2}). This is because of the particle-hole symmetry in our derivation. With this symmetry, the amplitude and phase fluctuations represent mutually orthogonal excitations\cite{Am0}, and hence, are decoupled\cite{symmetry}. Moreover, with the particle-hole symmetry, the disorder-induced local potential can not excite the amplitude fluctuation, since the charge-amplitude correlation vanishes according to recent symmetry analysis\cite{symmetry}.

For the excited phase fluctuation ${\bf p}_s$, the solved $\rho_{{\bf k}0}$ in clean case reads (refer to Appendix~\ref{app2})
\begin{equation}\label{rk0}
\rho_{{\bf k}0}=({\bf v}_{\bf k}\cdot{\bf p}_s)\frac{\Delta^2_{\bf k}}{E_{\bf k}}\partial_{E_{\bf k}}F_{\bf k}.  
\end{equation}
Then, from Eq.~(\ref{current}), the generated supercurrent is given by
\begin{equation}\label{current1}
{\bf j}=\frac{e{\bf p_s}}{m}\frac{k_F^2}{m}{\sum_{\bf k}}'\frac{\Delta^2_{\bf k}}{E_{\bf k}}\partial_{E_{\bf k}}{F_{\bf k}}=\frac{{e}n_s}{m}{\bf p}_s,  
\end{equation}
from which one obtains the superfluid density in Eq.~(\ref{SD}).

\subsubsection{Scattering of momentum relaxation}

For the scattering part, we focus on the momentum (current) relaxation. Then, according to Eq.~(\ref{current}), one only needs to keep the $\tau_0$ components of $\rho_{\bf k}$ and $\partial_{t}\rho_{\bf k}\big|_{\rm scat}$, and hence, Eq.~(\ref{scat1}) becomes
\begin{widetext}
\begin{eqnarray}
 \!\!\!\! \partial_{t}\rho_{\bf k}\Big|^{\rm mr}_{\rm scat}\!\!&=&\!\!-2n_i\pi\sum_{{\bf k'}\eta\eta'}|U^s_{\bf kk'}|^2{\rm Tr}(\tau_3\Gamma^{\eta}_{\bf k}\tau_3\Gamma^{\eta'}_{\bf k'})(\rho_{{\bf k}0}-\rho_{{\bf k}'0})\delta(E^{\eta}_{\bf k}-E^{\eta'}_{\bf k'})\nonumber\\
\!\!&=&\!\!-2n_i\pi\sum_{{\bf k'}}|U^s_{\bf kk'}|^2(\rho_{{\bf k}0}\!-\!\rho_{{\bf k}'0})[{e^-_{\bf kk'}}\delta(E^{+}_{\bf k}\!-\!E^{+}_{\bf k'})\!+\!{e^-_{\bf kk'}}\delta(E^{-}_{\bf k}\!-\!E^{-}_{\bf k'})\!+\!{e^+_{\bf kk'}}\delta(E^{+}_{\bf k}\!-\!E^{-}_{\bf k'})\!+\!{e^+_{\bf kk'}}\delta(E^{-}_{\bf k}\!-\!E^{+}_{\bf k'})],~~~    
\end{eqnarray}
\end{widetext}
where $e_{\bf kk'}^{\pm}=\frac{1}{2}(1\pm\frac{\Delta_{\bf k}\Delta_{\bf k'}-\xi_{\bf k}\xi_{\bf k'}}{E_{\bf k}E_{\bf k'}})$. It is noted that the term ${e^-_{\bf kk'}}(\rho_{{\bf k}0}\!-\!\rho_{{\bf k}'0})[\delta(E^{+}_{\bf k}\!-\!E^{+}_{\bf k'})\!+\!\delta(E^{-}_{\bf k}\!-\!E^{-}_{\bf k'})]$ vanishes around the Fermi surface. Consequently, one has
\begin{equation}\label{SCAT}
\partial_{t}\rho_{\bf k}\Big|^{\rm mr}_{\rm scat}\!=\!-\!{\sum_{{\bf k'}\eta=\pm}}'|M_{\bf kk'}|^2(\rho_{{\bf k}0}-\rho_{{\bf k}'0})\delta(E^{\eta}_{\bf k}-E^{-\eta}_{\bf k'}),  
\end{equation}
with $|M_{\bf kk'}|^2=2n_i\pi|U^s_{\bf kk'}|^2e^+_{\bf kk'}$. 

We next focus on the momentum-relaxation rate of the ${\bf k}$ particle in superfluid. In Eq.~(\ref{SCAT}), if ${\bf k}$ particle lies in the pairing region, one has ${\bf k}'$ particle lying in the unpairing region, as analyzed in Sec.~\ref{MDDsec}. Then, considering the situation with the drive effect from superconducting momentum ${\bf p}_s$, from Eq.~(\ref{rk0}), one has vanishing $\rho_{{\bf k}'0}$ and finite $\rho_{{\bf k}0}$. Consequently, Eq.~(\ref{SCAT}) becomes 
\begin{equation}
\partial_{t}\rho_{\bf k}\Big|^{\rm mr}_{\rm scat}\!=\!-\Gamma_{\bf k}\rho_{{\bf k}0},  
\end{equation}
with the momentum-relaxation rate of superfluid given by Eq.~(\ref{SCF}). For weak external probe that is related to the intrinsic character of system, one can use $\Gamma_{\bf k}$ to discuss the superconductivity phenomenon, with ${\bf p}_s$ in $\Gamma_{\bf k}$ arising from the phase fluctuation.

\subsubsection{Phase-coherence dynamics}
\label{PCDsec}

We next construct the macroscopic phase-coherence dynamics by deriving the equation of motion of the phase fluctuation. We neglect the mutual interaction between phase fluctuations by only keeping the linear order of phase fluctuation in its equation of motion. Then, in center-of-mass frequency and momentum space $[(t,{\bf R})\rightarrow(\omega,{\bf q})]$, substituting the solved $\rho_{{\bf k}2}$ into Eq.~(\ref{phase}),  the equation of motion of the phase fluctuation is written as (refer to Appendix~\ref{app2})
\begin{equation}
2i\omega({i\omega\delta\theta_{\bf q}}/{2}+\mu_{H})D+i{\bf q}\cdot{\bf p}_sn_s/m=-2i\omega{DU_{\bf q}}.\label{PE1}  
\end{equation}
As seen from the left-hand side of above equation, without the Hartree field $\mu_H$, one immediately finds a gapless energy spectrum of the phase fluctuation with the linear dispersion, in consistency with the Goldstone theorem\cite{Gm1,Gm2} by the spontaneous breaking of continuous $U(1)$ symmetry in superconductors\cite{gi0}. 

Particularly, substituting the solution of $\rho_{{\bf k}3}$ into Eq.~(\ref{density}), the density fluctuation reads (refer to Appendix~\ref{app2})
\begin{equation}
\delta{n}=-2D\mu_{\rm eff}.\label{density1}
\end{equation}
Then, it is noted that from the expressions of the density [Eq.~(\ref{density1})] and current [Eq.~(\ref{current1})], the equation of motion of the phase fluctuation in Eq.~(\ref{PE1}) is a direct consequence of the charge conservation $\partial_t\delta{n}+{\bm \nabla_{\bf R}}\cdot{\bf j}=0$. The charge conservation in the gauge-invariant kinetic theory is natural\cite{GIKE4}, as it has been proved long time ago by Nambu through the generalized Ward identity that the gauge invariance in the superconducting states is equivalent to the charge conservation~\cite{gi0}.

From Eq.~(\ref{density1}), the Hartree field [Eq.~(\ref{HF})] is therefore written as $\mu_H=-2V_{q}D\mu_{\rm eff}$. Then, Eq.~(\ref{PE1}) becomes
\begin{equation}
D_{q}\Big(\omega_p^2+\frac{n_sq^2}{2Dm}-\omega^2\Big)\frac{\delta\theta_{\bf q}}{2}=-i\omega{U_{\bf q}}D_{q},  \label{PEF}
\end{equation}
with $D_q=D/(1+2DV_q)$. As seen from the left-hand side of above equation, when the long-range Coulomb interaction is included, the original energy spectrum of the phase fluctuation is altered as {\small $\omega_N=\sqrt{\omega_p^2+n_sq^2/(2Dm)}$}. The right-hand side of above equation represents the source term from impurity potential. It is noted that additional source terms emerge on the right-hand side of Eq.~(\ref{PEF}) if other quantum disorder/fluctuation effects that generate electric potential or couple to phase fluctuation are considered, so $U_q$ here can represent an effective electric potential by related quantum disorder/fluctuation effects. The source term from the disorder-induced local electric potential excites the macroscopic inhomogeneous phase fluctuation through the Josephson effect, since without the long-range Coulomb interaction and kinetic term {${n_sq^2}/{(2Dm)}$}, Eq.~(\ref{PEF}) reduces to {$\partial_t\delta\theta/2=-U(R)$} in space-time coordinate, same as the Josephson effect\cite{Josephson}. 

From the equation of motion in Eq.~(\ref{PEF}), in principle, one can directly solve the generated phase fluctuation $\delta\theta(R)$ from disorder effect, and then, calculate the induced superconducting momentum ${\bf p}_s(R)=\nabla_{\bf R}\delta\theta(R)/2$ and hence Doppler shift, in order to further formulate the influence of the phase fluctuation on electronic fluids at each ${\bf R}$. Nevertheless, for the experimental detections that usually are long-wave measurement, one only needs to consider the long-wave component of the Doppler shift effect, which leads to a homogeneous influence on the electronic fluids. In this circumstance, we apply a simplified way by using the equation of motion in Eq.~(\ref{PEF}) to construct the action of the phase fluctuation:
\begin{equation}\label{action}
S\!=\!\int{dt}\!\sum_{\bf q}D_q\Big(\Big|\frac{\partial_t\theta}{2}\Big|^2\!-\!\omega^2_N\Big|\frac{\theta_{\bf q}}{2}\Big|^2\!+\!U^*_{\bf q}\frac{\partial_t\theta_{\bf q}}{2}\!+\!U_{\bf q}\frac{\partial_t\theta_{\bf q}^*}{2}\Big).  
\end{equation}
Then, on one hand, one can directly use above action to derive the expected value of the generated superconducting momentum from disorder effect. On the other hand, by mapping the action in Eq.~(\ref{action}) into the imaginary-time one $\mathcal{S}$, the expected value of the generated superconducting momentum from the thermal excitation can also be determined.

Consequently, considering the anisotropy in $d$-wave superconductors, the generated superconducting momentum by phase fluctuation along ${\bf e}_{\phi}$ direction reads
\begin{eqnarray}
  (p_s^{\phi})^2\!\!&\!=\!&\!\!\sum_qq^2\Big[\Big\langle\Big|\frac{\delta\theta^*_{q{\bf e}_{\phi}}(t+0^+)}{2}\frac{\delta\theta_{q{\bf e}_{\phi}}(t)}{2}e^{iS}\Big|\Big\rangle \nonumber\\
    &&\mbox{}+\Big\langle\Big|\frac{\delta\theta^*_{q{\bf e}_{\phi}}(\tau)}{2}\frac{\delta\theta_{q{\bf e}_{\phi}}(\tau)}{2}e^{-\mathcal{S}}\Big|\Big\rangle\Big]  \nonumber\\
  &\!=\!&\!\!\sum_qq^2\Big[\frac{1}{Z_0}{\int}D\theta{D\theta^*}\frac{\delta\theta^*_{q{\bf e}_{\phi}}(t+0^+)}{2}\frac{\delta\theta_{q{\bf e}_{\phi}}(t)}{2}e^{iS} \nonumber\\
&&\mbox{}+\frac{1}{\mathcal{Z}_0}{\int}D\theta{D\theta^*}\frac{\delta\theta^*_{q{\bf e}_{\phi}}(\tau)}{2}\frac{\delta\theta_{q{\bf e}_{\phi}}(\tau)}{2}e^{-\mathcal{S}}\Big], \label{GFM}
\end{eqnarray}
with $Z_0$ and $\mathcal{Z}_0$ standing for the corresponding partition functions. Following the standard generating functional method\cite{FT} to handle above equation (refer to Appendix~\ref{app3}), one has 
\begin{equation}
  (p_s^{\phi})^2\!=\!\frac{i}{C}\sum_{q\omega}\frac{q^2\omega^2U_{q{\bf e}_{\phi}}U_{-q{\bf e}_{\phi}}}{(\omega^2\!-\!\omega_N^2+i0^+)^2}\!-\!\frac{1}{\beta}\sum_{q\omega_n}\frac{q^2}{D_q}\frac{1}{(i\omega_n)^2\!-\!\omega_N^2}, \label{GFMM}
\end{equation} 
where $\omega_n=2n\pi{T}$ denotes the Matsubara frequency and $\beta=1/(k_BT)$ with $k_B$ being the Boltzmann constant. Then, after the frequency and Matsubara-frequency summations, Eq.~(\ref{PS1}) can be directly obtained.

\subsection{Path integral approach}

In this part, in order to confirm our derivation from the GIKE approach, we use the path integral approach\cite{pi1,pi4} to derive the macroscopic phase-coherence dynamics and microscopic electronic-fluid dynamics. We start with the generalized action of superconductors:
\begin{eqnarray}
 && S[\phi,\phi^*]=\sum_{s=\uparrow,\downarrow}\int{dx}\psi^*_s(x)[i\partial_t-\xi_{\hat {\bf p}}-U(x)]\psi_s(x)\nonumber\\
  &&\mbox{}-\frac{1}{2}\sum_{ss'}\int{dxdx'}V(x-x')\psi_s^*(x)\psi_{s'}^*(x')\psi_{s'}(x')\psi_s(x) \nonumber\\
   &&\mbox{}+\int{dxdx'}g(x-x')\psi^*_{\uparrow}(x)\psi^*_{\downarrow}(x')\psi_{\downarrow}(x')\psi_{\uparrow}(x).
\end{eqnarray}
All symbols used here are same as the previous ones. Applying the Hubbard-Stratonovich transformation, the above action becomes
\begin{eqnarray}
  S[\psi,\psi^*]\!\!&=&\!\!\!\!\sum_{s=\uparrow,\downarrow}\int{dx}\psi^*_s(x)[i\partial_t\!-\!\xi_{\hat {\bf p}}\!-\!U(x)\!-\!\mu_H(x)]\psi_s(x)\nonumber\\
  &&\mbox{}-\int{dxdx'}\psi^*(x){\hat \Delta}(x,x')\psi(x')\nonumber\\
  &&\mbox{}-\!\int{d^4R}\frac{|\Delta|^2}{g}+\frac{1}{2}\sum_{\omega,{\bf q}}\frac{|\mu_H(q)|^2}{V_q}.
\end{eqnarray}
in which we have substituted the pairing potential $g_{\bf kk'}=g\cos(\zeta\theta_{\bf k}+\alpha)\cos(\zeta\theta_{\bf k'}+\alpha)$ to give rise to {\small $\Delta_{{\bf k}}=\Delta\cos(\zeta\theta_{\bf k}+\alpha)$}, with $\zeta$ being the orbital angular momentum of the pairing symmetry. By further using the unitary transformation in Eq.~(\ref{utrp}) to effectively remove the phase fluctuation from the order parameter, one has
\begin{eqnarray}
&&S[\psi,\psi^*]\!=\!\int{dxdx'}\psi^*(x)[G^{-1}_0(x,x')\!-\!\Sigma(R)\tau_3]\psi(x')\nonumber\\
  &&\mbox{}-\!V_f\!\int{d^4R}\Sigma(R)\!-\!\int{d^4R}\frac{|\Delta|^2}{g}\!+\!\frac{1}{2}\sum_{\omega,{\bf q}}\frac{|\mu_H(q)|^2}{V_q},
\end{eqnarray}
where $V_f=\sum_{\bf k}1$ arises from the anti-commutation of the Fermi field; the Green function is written as
\begin{equation}\label{GF0}
  G^{-1}(x,x')=[i\partial_t-{\bf p}_s\cdot{\hat {\bf p}}/m-\xi_{\hat {\bf p}}\tau_3]\delta(x-x')-|\Delta(x,x')|\tau_1,  
\end{equation}
and the self-energy reads
\begin{equation}\label{SEE}
  \Sigma(R)=\mu_{\rm eff}(R)+\frac{p_s^2}{2m}.  
\end{equation}
It is noted that in the previous works\cite{pi1,pi4}, the Doppler-shift term ${\bf p}_s\cdot{\hat {\bf p}}/m$ is placed into the self-energy $\Sigma(R)$ and then treated as small quantity to take perturbation expansion. This approximation holds only if $|{\bf p}_s\cdot{{\bf v}_{\bf k}}|<\Delta_{\bf k}$, while this condition is usually satisfied in conventional $s$-wave superconductors with inactive phase fluctuation or weak external electromagnetic field. In the present work, considering the anisotropy and strong phase fluctuation in $d$-wave cuprate superconductors, we sublate this approximation by taking the Doppler-shift term into the Green function (i.e., quasiparticle energy spectra).

Then, after the standard integration over the Fermi field, one has
\begin{eqnarray}
S&=&{\rm {\bar Tr}}\ln{[G_0^{-1}-\Sigma\tau_3]}\nonumber\\
 &&\mbox{}-\!V_f\!\int{d^4R}\Sigma(R)-\!\int{d^4R}\frac{|\Delta|^2}{g}+\frac{1}{2}\sum_{\omega,{\bf q}}\frac{|\mu_H(q)|^2}{V_q}\nonumber\\
&=&{\rm {\bar Tr}}\ln{G_0^{-1}}-\frac{1}{n}\sum_{n=1}^{\infty}{\rm {\bar Tr}}[(G_0\Sigma\tau_3)^n]\nonumber\\
 &&\mbox{}-\!V_f\!\int{d^4R}\Sigma(R)-\!\int{d^4R}\frac{|\Delta|^2}{g}+\frac{1}{2}\sum_{\omega,{\bf q}}\frac{|\mu_H(q)|^2}{V_q}\nonumber\\
&\approx&\int{d^4{R}}{\sum_{p_n,{\bf k}}}\ln[(ip_n-E_{\bf k}^+)(ip_n-E_{\bf k}^-)]\nonumber\\
  &&\mbox{}-\!Q_1\int{d^4R}\Sigma(R)\!+\!Q_2\int{d^4R}[\Sigma(R)]^2\nonumber\\
  &&\mbox{}-\!\int{d^4R}\frac{|\Delta|^2}{g}+\frac{1}{2}\sum_{\omega,{\bf q}}\frac{|\mu_H(q)|^2}{V_q}, \label{S1}
\end{eqnarray}
in which we have neglected the mutual interaction between phase fluctuations by only keeping the lowest two orders (i.e., $n=1$ and $n=2$). Here, $p_n=(2n+1)\pi{T}$ denotes the Matsubara frequency and the coefficients read (refer to Appendix~\ref{app4})
\begin{eqnarray}
   Q_1\!\!\!&=&\!\!\!V_f\!+\!\sum_p{\rm Tr}[G_0(p)\tau_3]=\!-\frac{k_F^2}{m}{\sum_{\bf k}}'\partial_{\xi_{\bf k}}\big(\xi_{\bf k}F_{\bf k}\big),~~~~\label{Q1} \\
  Q_2\!\!\!&=&\!\!\!-\frac{1}{2}\sum_p{\rm Tr}[G_0(p)\tau_3G_0(p)\tau_3]=\!-\!{\sum_{\bf k}}'\partial_{\xi_{\bf k}}\big(\xi_{\bf k}F_{\bf k}\big).~~~~\label{NDE} 
\end{eqnarray}
Here, $p=(ip_n,{\bf k})$ and the Green function $G(p)$ from Eq.~(\ref{GF0}) is given by
\begin{equation}\label{GFP}
G_0(p)=\frac{ip_n\tau_0-{\bf p}_s\cdot{\bf v_k}\tau_0+\xi_{\bf k}\tau_3+\Delta_{\bf k}\tau_1}{(ip_n-E_{\bf k}^+)(ip_n-E_{\bf k}^-)}.  
\end{equation}
As {\small $\mu_{H}(q=0)=V_{q=0}\delta{n_{q=0}}=0$}, it can be easily demonstrated that $\int{d^4R}\Sigma(R)=\int{d^4R}[U(R)+p^2_s/(2m)]$. Then, one obtains the effective action of superconductors:
\begin{eqnarray}
  S_{\rm eff}&=&\int{d^4{R}}\Big\{{\sum_{p_n,{\bf k}}}\ln[(ip_n-E_{\bf k}^+)(ip_n-E_{\bf k}^-)]\!-\!\frac{|\Delta|^2}{g}\nonumber\\
  &&\mbox{}\!+\!Q_2\Big(\mu_{\rm eff}^2\!-\!\frac{k_F^2p_s^2}{2m^2}\Big)\Big\}\!+\!\frac{1}{2}\sum_{\omega,{\bf q}}\frac{|\mu_H(q)|^2}{V_q}.  \label{S2}
\end{eqnarray}
We next handle the Hartree field. Through the integration over the Hartree field in Eq.~(\ref{S2}), one gets
\begin{eqnarray}
&&{\bar S}_{\rm eff}=\int{d^4{R}}{\sum_{p_n,{\bf k}}}\Big\{\ln[(ip_n-E_{\bf k}^+)(ip_n-E_{\bf k}^-)]\!-\!\frac{|\Delta|^2}{g}\Big\}\nonumber\\
&&\mbox{}+\!\!\int{dt}\sum_{\bf q}\Big[\frac{Q_2}{1\!+\!2Q_2V_q}\Big(\frac{\partial_{t}\delta\theta_{\bf q}}{2}\!+\!U_{\bf q}\Big)^2\!-\!\frac{Q_2k_F^2p_s^2}{2m^2}\Big].~~~~\label{actionF}  
\end{eqnarray}
From the action above, we prove in the following that the previous gap equation [Eq.~(\ref{GE1})], superfluid density [Eq.~(\ref{SD})] as well as the equation of motion of the phase fluctuation [Eq.~(\ref{PEF})] derived from the GIKE approach can all be recovered. 

Specifically, through the variation $\delta{{\bar S}_{\rm eff}}=0$ with respect to the gap, one has
\begin{eqnarray}
  \Delta&=&-g{\sum_{\bf k}}'\Delta\cos^2(\zeta\theta_{\bf k}+\alpha)\sum_{p_n}\frac{1}{(ip_n-E_{\bf k}^+)(ip_n-E_{\bf k}^-)}\nonumber\\
  &=&-g{\sum}'_{\bf k}\Delta_{\bf k}\cos(\zeta\theta_{\bf k}+\alpha)F_{\bf k},
\end{eqnarray}
which with $g_{\bf kk'}=g\cos(\zeta\theta_{\bf k}+\alpha)\cos(\zeta\theta_{\bf k'}+\alpha)$, is exactly same as the previous gap equation in Eq.~(\ref{GE1}). The supercurrent is given by
\begin{eqnarray}
  {\bf j}&=&-e{\partial_{{\bf p}_s}{{\bar S}_{\rm eff}}}\!=\!\frac{eQ_2k_F^2{\bf p_s}}{m^2}\!+\!{\sum_{p_n{\bf k}}}\frac{2e{\bf k}(ip_n\!-\!{\bf k}\!\cdot\!{\bf p}_s/m)}{m(ip_n\!-\!E_{\bf k}^+)(ip_n\!-\!E_{\bf k}^-)}\nonumber\\
  &=&-\frac{ek_F^2{\bf p_s}}{m^2}{\sum_{\bf k}}'\Big\{\partial_{\xi_{\bf k}}(\xi_{\bf k}F_{\bf k})-\partial_{E_{\bf k}}\Big[\frac{f(E_{\bf k}^+)-f(E_{\bf k}^-)}{2}\Big]\Big\}\nonumber\\
  &=&\frac{e{\bf p_s}}{m}\frac{k_F^2}{m}{\sum_{\bf k}}'\frac{\Delta_{\bf k}^2}{E_{\bf k}}\partial_{E_{\bf k}}F_{\bf k}=\frac{en_s{\bf p_s}}{m}.
\end{eqnarray}
Then, the previous supercurrent in Eq.~(\ref{current1}) and hence superfluid density in Eq.~(\ref{SD}) are recovered.

Through the variation $\delta{{\bar S}_{\rm eff}}=0$ with respect to the phase fluctuation, one obtains
\begin{eqnarray}  
  0&=&\partial_{t}\Big[\frac{\partial{\bar S}_{\rm eff}}{\partial(\partial_t\delta\theta_{\bf q}/2)}\Big]-i{\bf q}\cdot\Big(\frac{\partial{\bar S}_{\rm eff}}{\partial{\bf p}_s}\Big)\nonumber\\
  &=&\frac{2Q_2}{1+2Q_2V_q}\partial_t\Big(\frac{\partial_{t}\delta\theta_{\bf q}}{2}+U_{\bf q}\Big)-{i{\bf q}\cdot{\bf p}_s}\frac{n_s}{m}\nonumber\\
&=&D_{q}\Big(\frac{q^2n_s}{2Dm}+\frac{q^2V_qn_s}{m}+\partial^2_t\Big)\frac{\delta\theta_{\bf q}}{2}+D_{q}\partial_tU_{\bf q},  
\end{eqnarray}
in which we have used $Q_2=-\sum_{\bf k}'\partial_{\xi_{\bf k}}(\xi_{\bf k}F_{\bf k})\approx{D}$. Then, the previous equation of motion of the phase fluctuation in Eq.~(\ref{PEF}) is recovered.

Consequently, the previous gap equation [Eq.~(\ref{GE1})], superfluid density [Eq.~(\ref{SD})] as well as the equation of motion of the phase fluctuation [Eq.~(\ref{PEF})] derived from the GIKE approach can all be recovered by the path-integral approach. Whereas the microscopic scattering of electronic fluids (i.e., microscopic momentum-relaxation rate in superfluid) is hard to handle within the path-integral approach. The conventional Kubo diagrammatic formalism is also difficult to track the microscopic scattering, as the inevitable calculation of the vertex correction becomes hard to deal with in superconductors\cite{G1}, especially considering the anisotropy and significant phase fluctuation (i.e., Doppler shift) in $d$-wave case. Thus, the GIKE approach in fact provides an efficient way to deal with the microscopic scattering in superconductors for investigating the superconductivity (zero-resistance) phenomenon.

\section{Application to disordered $s$-wave superconductors}

\begin{figure}[htb]
   {\includegraphics[width=9.0cm]{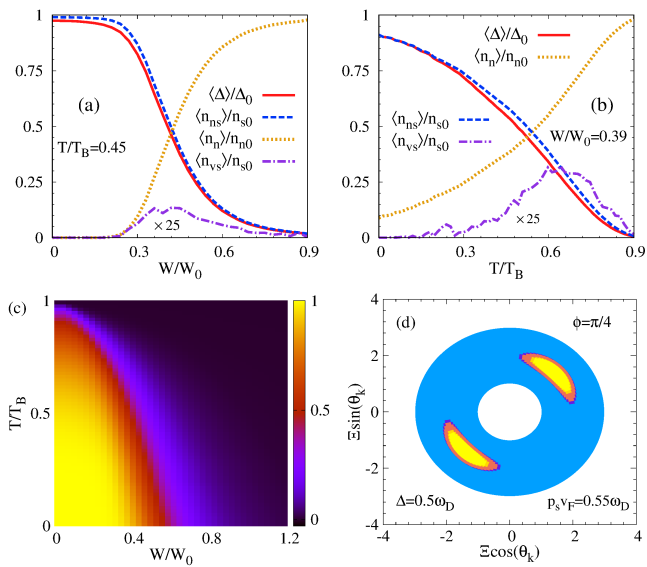}}
   \caption{(Color online) (a) Anderson-disorder strength and (b) temperature dependence of gap $\langle\Delta\rangle$ and densities of the non-viscous superfluid $\langle{n_{\rm ns}}\rangle$, viscous superfluid $\langle{n_{\rm vs}}\rangle$ and normal fluid $\langle{n_{n}}\rangle$ as well as (c) phase diagram of $\langle\Delta\rangle$ from the full and self-consistent numerical simulation in disordered $s$-wave superconductors. The results of $\langle{n_{\rm vs}}\rangle$ in figures (a) and (b) are multiplied by $25$ for the visibility. Note that the small deviation of $\Delta/\Delta_0$ from one at $W=0$ in figure (a) originates from the finite temperature. (d) Schematic showing the division in the momentum space at ${\bf p}_s=0.55\omega_D/v_F{\bf e}_{\phi=\pi/4}$ and $\Delta=0.5\omega_D$ in $s$-wave superconductors. In the figure, $T_{B}$ denotes the critical temperature where gap vanishes in the absence of the phase fluctuation, i.e., BCS critical temperature of the two-dimensional $s$-wave superconductors; $n_{s0}$ represents the superfluid density at zero temperature in the absence of the phase fluctuation, and $n_{n0}$ denotes the normal-fluid density above $T^{\rm os}$; $W_0=0.5\Delta_0$. }  
 \label{figyw5}
\end{figure}

Finally, considering the recent experimental progress of realizing the atomically thin superconductors\cite{atsc01,atsc02,atsc03,atsc04,atsc1} and disordered superconducting films\cite{DSTF1,DSTF2,DSTF3,DSTF4} with $s$-wave gap, we briefly investigate the two-dimensional disordered $s$-wave superconductors. In this circumstance, the phase fluctuation retains gapless energy spectrum even after considering the long-range Coulomb interaction, and hence, is active. We therefore apply the developed dual dynamics with gap $\Delta_{\bf k}=\Delta$ and pairing potential $g_{\bf kk'}=g$. By using the similar numerical simulation as the $d$-wave case in Sec.~\ref{sec-NS}, the obtained results in disordered $s$-wave superconductors are plotted in Fig.~\ref{figyw5}.

Actually, based on the three-fluid model, by applying the analysis as the $d$-wave case in Sec.~\ref{sec-TD}, one also expects a similar behavior between the non-viscous superfluid density $\langle{n_{\rm ns}}\rangle$ and gap $\langle\Delta\rangle$ and a compensatory behavior between $\langle{n_{\rm ns}}\rangle$ and normal-fluid density $\langle{n_{n}}\rangle$ in temperature (Anderson-disorder strength) dependence of disordered $s$-wave superconductors, as confirmed by numerical results in Fig.~\ref{figyw5}(b) [Fig.~\ref{figyw5}(a)].

The difference from the $d$-wave case in disordered $s$-wave superconductors includes three aspects. Firstly, as mentioned in Sec.~\ref{sec-TD}, differing from the $d$-wave case where nonzero normal fluid always emerges around nodal points, the emergence of the normal fluid in $s$-wave superconductors requires ${p}_s>\Delta/v_F$, leading to a threshold to realize normal fluid\cite{GIKE1}. Therefore, at small phase fluctuation with $p_s<\Delta/v_F$, no normal fluid and hence viscous superfluid are generated  in $s$-wave superconductors, while the gap and non-viscous superfluid density are free from the influence of the phase fluctuation (i.e., increase of $W$), as shown in Fig.~\ref{figyw5}(a) at $W<0.24$. 

Secondly, with the increase of phase fluctuation (by enhancing $W$) at $p_s>\Delta/v_F$, the gap $\langle\Delta\rangle$ in disordered $s$-wave superconductors [red solid curve in Fig.~\ref{figyw5}(a)] shows a much faster decrease than that in $d$-wave case [Fig.~\ref{figyw2}(i)]. This implies that the $s$-wave gap with the higher rotational symmetry is much more fragile against the phase fluctuation. At large phase fluctuation, as shown in Fig.~\ref{figyw5}(b), the temperature dependence of $\langle\Delta\rangle$ (red solid curve) deviates from the BCS-like behavior and exhibits a faster decrease with temperature, and a nonzero fraction of normal fluid emerges even at zero temperature. These two behaviors provide a scheme for experimental detection to justify the existence of phase fluctuation in disordered $s$-wave superconductors. 

Furthermore, as seen from Figs.~\ref{figyw5}(a) and~(b), in disordered $s$-wave superconductors, the viscous superfluid density $\langle{n_{\rm vs}}\rangle$ is marginal, and hence, the normalized non-viscous superfluid density $\langle{n_{\rm ns}}\rangle/n_{s0}$ (blue dashed curve) nearly coincides with the normalized gap $\langle\Delta\rangle/\Delta_0$ (red solid curve), in sharp contrast to the results in $d$-wave case. This arises from the isotropy of the $s$-wave gap. Specifically, in disordered $s$-wave superconductors, as shown by the schematic illustration in Fig.~\ref{figyw5}(d), at phase fluctuation $p_s>\Delta/v_F$, even the normal fluid (yellow regions) is excited, the viscous pairing (orange) region that can experience the friction with normal fluid [i.e., satisfying the energy conservation of momentum-relaxation rate in Eq.~(\ref{SCF})] lies around the unpairing one and is very small, in sharp contrast to the large fraction in anisotropic $d$-wave case at the same condition [Figs.~\ref{figyw1}(b) and~(f)]. Moreover, the anomalous correlation $F_{\bf k}$ adjacent to the unpairing region is very small. Therefore, due to the small area and weak $F_{\bf k}$, the viscous pairing region in disordered $s$-wave superconductors makes a neglectable contribution in the gap equation [Eq.~(\ref{GE1})] and leads to a marginal $n_{\rm vs}$ [Eq.~(\ref{vsd})]. With the marginal $\langle{n_{\rm vs}}\rangle$, the separation between $T_c$ and $T^{\rm os}$ is remarkably small ($0.01T_B$ at most) in our simplified model.

\section{Summary}

In summary, by using the GIKE approach\cite{GIKE1,GIKE2,GIKE3,GIKE4}, we construct the coupled dual dynamics of macroscopic phase coherence and microscopic electronic fluids (consisting of normal fluid and superfluid) in cuprate superconductors. The macroscopic phase-coherence dynamics is developed by deriving the equation of motion of the phase fluctuation, in which both disorder and long-range Coulomb interaction effects are considered. We show that the phase fluctuation in cuprate superconductors retains gapless energy spectrum thanks to the layered structures\cite{cuprate4,cuprate5}, and hence, is active, differing from the conventional bulk superconductors with inactive phase fluctuation due to Anderson-Higgs mechanism\cite{AHM,AK,Am0,Ba0}. Moreover, the superfluid density determines the phase stiffness in the phase-coherence dynamics, in consistency with the previous understanding in the literature\cite{phase1,phase2,phase3}. The microscopic electronic-fluid dynamics includes two parts: the anomalous correlation to determine gap and superfluid density; the microscopic scattering of the electronic fluids that is essential for studying the transport property and hence superconductivity. It is shown that both anomalous correlation and microscopic scattering in the electronic-fluid dynamics are affected by the phase fluctuation as it drives the Doppler shift in quasiparticle energy spectra\cite{FF4,FF5,FF6,GIKE1} by generating a superconducting momentum\cite{gi0,Ba0,G1}.

Based on the developed dual dynamics, we present theoretical descriptions of the separation between $T_c$ and $T^{\rm os}$ as well as the emerged normal fluid below $T_c$ in cuprate superconductors. We find that the key origin of both phenomena comes from the quantum effect of disorder, which excites the macroscopic inhomogeneous phase fluctuation through Josephson effect.\cite{Josephson} With this excited phase fluctuation, we prove that the unpairing region with vanishing anomalous correlation\cite{FF1,FF2,FF7,FF8,FF9,GIKE1}, i.e., normal fluid, always emerges around nodal points in $d$-wave cuprate superconductors even at zero temperature, in consistency with the experimentally observed substantial fraction of normal state at low temperature\cite{NSB1,NSB2}. Whereas according to the microscopic scattering, the pairing region, i.e., superfluid, is divided into two parts: viscous and non-viscous ones. Particles in the viscous pairing region experience a finite momentum relaxation, due to the scattering with the ones in unpairing region, behaving like the friction between the superfluid and normal fluid. The non-viscous superfluid is free from the momentum relaxation scattering. Therefore,  in addition to conventional non-viscous superfluid, there also exist normal fluid and viscous superfluid at small phase fluctuation in cuprate superconductors, similar to the three-fluid model proposed in our previous work\cite{GIKE1} in conventional superconductors which is caused by external electromagnetic field. An experimental scheme to distinguish the densities of these three electronic fluids is proposed. 

We further demonstrate that by increasing the temperature in cuprate superconductors, the suppressed gap and hence superfluid density weaken the phase stiffness, enhancing the phase fluctuation. Once the phase fluctuation (i.e., temperature) exceeds the critical point, the non-viscous superfluid vanishes, leaving only normal fluid and viscous superfluid. The system then enters the pseudogap state with nonzero resistivity and finite gap due to the significant phase fluctuation. It is noted that in this circumstance, the viscous superfluid matches the description of the incoherent preformed Cooper pairs, as they both contribute to gap but experience the scattering. Whereas the existing normal fluid in our description implies the existence of normal particles in pseudogap state, which has been overlooked in previous preformed Cooper-pair model to describe pseudogap state\cite{phase2,PCP1,PCP2,PCP3,PCP4,PCP5}. The viscous superfluid starts to shrink by further increasing temperature in pseudogap state, until vanishes at $T^{\rm os}$ where gap is eventually destroyed.

To confirm the derivation from the GIKE approach, we also apply the path-integral approach to recover the equation of motion of the phase fluctuation as well as anomalous correlation, gap equation and superfluid density in the presence of the superconducting momentum. Holding the pairing potential fixed, a self-consistent numerical simulation by applying Anderson disorder is also addressed, to verify our theoretical description. Then, both the separation between $T_c$ and $T^{\rm os}$ as well as the emerged normal fluid below $T_c$ are confirmed. Particularly, $T^{\rm os}$ tracks $T_c$ closely in our numerical results, and lies well below the experimental pseudogap temperature $T^*$, in consistency with the current experimental findings and understanding in the literature\cite{TM2,TM3,TM4,SH,PC1,NF1,AF1,AF2,AF3,AF4,PPS1,PPS2,Infrad1,Infrad2,Infrad3,MS1,MS2,MS3,THZ1,THZ2,THZ3,THZ4,DHM3}. This is because that once in the pseudogap state, due to the significant phase fluctuation and hence Doppler shift, the gap can not survive far above $T_c$.   

Consequently, when the pairing potential is determined, the coupled dual dynamics in the present work provides an efficient and simplified approach to formulate the dephasing process of macroscopic superconducting phase coherence with the increase of temperature, as well as the influence of this dephasing on microscopic electronic fluids (including gap, densities of superfluid and normal fluid, and in particular, the transport property to determine superconducting transition temperature $T_c$). This theory distinguishingly takes the Anderson-disorder approach and in particular, impurity scattering treatment to discuss the disorder effects on macroscopic phase coherence and microscopic electronic fluids, respectively. The Anderson-disorder approach that calculates the converged quantities by averaging over numerous random impurity configuration is applied in order to characterize the quantum disorder/fluctuation effects from the impurity as well as possible couplings to charge-density wave\cite{cuprate5,CDW0,CDW1,CDW2,CDW3}, spin-density wave\cite{cuprate5,SDW1,SDW2,SDW3,SDW4}, electronic nematicity\cite{cuprate5,Nernst6,NO1,NO2} and/or theoretically proposed spinon-vortices excitation\cite{VP1,VP2,VP3}.  Whereas the microscopic scattering, as the essential transport property for studying the superconductivity (zero-resistance) phenomenon, is still absent in the literature, since it is hard to handle in path-integral approach and conventional Kubo diagrammatic formalism. But the GIKE approach straightly tackles this crucial problem. Further determining the pairing potential and Anderson-disorder strength at different dopings requires the microscopic pairing mechanism, which still remains an open question in the literature and is beyond the scope of present work.

Considering the recent experimental progress of realizing the atomically thin superconductors\cite{atsc01,atsc02,atsc03,atsc04,atsc1} and disordered superconducting films\cite{DSTF1,DSTF2,DSTF3,DSTF4} with the $s$-wave gap, we show that the developed dual dynamics can also be applied similarly to the low-dimensional disordered $s$-wave superconductors, which exhibits an active phase fluctuation due to the gapless energy spectrum. Differing from the $d$-wave case where nonzero normal fluid always emerges around nodal points, there exists a threshold for phase fluctuation to induce normal fluid in isotropic $s$-wave superconductors\cite{GIKE1}. Consequently, at small phase fluctuation below threshold, no normal fluid and hence viscous superfluid are generated in $s$-wave superconductors, while the gap and non-viscous superfluid density are free from the influence of the phase fluctuation. But for large phase fluctuation above threshold, we find that the $s$-wave  gap with higher rotational symmetry is much more fragile against the phase fluctuation. In this situation, the temperature dependence of gap deviates from the BCS-like behavior and exhibits a faster decrease with temperature, and a nonzero fraction of normal fluid emerges even at zero temperature. Nevertheless, due to the isotropy of $s$-wave gap, the viscous superfluid density is marginal in our simplified model, leading to a remarkably small separation between $T_c$ and $T^{\rm os}$.

\section{Discussion}

Finally, we would like to discuss recent theoretical and experimental progresses in cuprate superconductors. 

{\em Role of bandstructure.---}The present work approximately takes the parabolic spectrum. Whereas recently, from the experimental angle-resolved photo-emission measurement, He {\em et al.} reported an important role of the flat energy dispersion near antinodal point in determining $T_c$ (i.e., phase coherence) in cuprate superconductors\cite{FB}. Here, based on our coupled dual dynamics, we briefly discuss this effect. Specifically, around Fermi surface, with the flat (i.e., $v_{\bf k}\approx0$) and steep (i.e., large $v_{\bf k}$) energy dispersions near antinodal and nodal points, respectively, a large unpairing region ($|{\bf v}_{\bf k}\cdot{\bf p}_s|>E_{\bf k}$) around nodal points can be achieved easily in the presence of significant phase fluctuation. Whereas the region around antinodal point is the pairing one ($|{\bf v}_{\bf k}\cdot{\bf p}_s|<E_{\bf k}$), but can become viscous through the friction with the large unpairing region. Particularly, since the Doppler shift ${\bf v}_{\bf k}\cdot{\bf p}_s\approx0$ nearby, this viscous pairing region can survive against very large phase fluctuation. Consequently, a viscous superfluid that contributes to gap can exist far above $T_c$, enlarging the separation between $T_c$ and $T^{\rm os}$.
Moreover, following the similar analysis, one can also expect a larger separation between $T_c$ and $T^{\rm os}$ in low-dimensional disordered $s$-wave superconductors with complex Fermi surface. The numerical simulation of this scheme in cuprate or disordered $s$-wave superconductors requires a specific calculation of the bandstructure, which is beyond the scope of the present work. 

{\em Disorder treatment.---}Recently, Lee-Hone {\em et al.}\cite{DS1,DS2} calculated the self-energy from the impurity and then self-consistently formulated the renormalizations of the gap and superfluid density in $d$-wave cuprate superconductors. A finite disorder influence is obtained in their results, differing from the vanishing one in conventional $s$-wave superconductors (Anderson theorem)\cite{AT1,AT2,AT3}. It is noted that this disorder treatment which takes the random phase approximation focuses on the diffusive-motion influence on the equilibrium gap $\Delta_{\bf k}$. The disorder effect on the phase coherence or amplitude fluctuation as well as the microscopic scattering of the nonequilibrium response are beyond their equilibrium calculation that carried out with the translational symmetry. 

{\em Amplitude fluctuation.---}Another issue concerns the amplitude fluctuation of the order parameter. As mentioned in Sec.~\ref{sec-GESD}, in the leading order of the disorder effect where the particle-hole symmetry is present, one finds a vanishing amplitude fluctuation. But the phase fluctuation is generated by Josephson effect and then influences the electronic fluids by inducing a superconducting momentum, as the present work addressed. Whereas in higher-order contribution with the broken particle-hole symmetry, the disorder-induced local potential can directly excite the amplitude fluctuation, as the charge-amplitude correlation becomes nonzero\cite{symmetry}. 

Recently, by using the tight-binding model in real space and applying the Anderson disorder, Li {\em et al.} performed a self-consistent numerical calculation to solve the wave function and hence gap\cite{disorder}. A drastic amplitude fluctuation emerges in their results by disorder effect, leading to granular superconducting islands where the gap is destroyed in strong-disorder regions. They attributed this formation to the pairing-breaking effect on anisotropic gap from disorder, and then, suggested that the strong phase fluctuation emerges in regions with small gap (i.e., superfluid stiffness) as a consequence. This approach clearly handles the high-order disorder effect (i.e., generation of amplitude fluctuation) well, but the leading contribution via Josephson effect [i.e., $\partial_t\delta\theta/2=-U(R)$] to excite the phase fluctuation is not involved in this stationary calculation. Particularly, in this situation with strong disorder, the phase fluctuation is likely to destroy the global gap before the drastic amplitude fluctuation emerges. Even the global gap survives, the influence of the phase fluctuation on such state with formation of granular superconducting islands can not be overlooked.

\begin{acknowledgments}
The authors acknowledge financial support from
the National Natural Science Foundation of 
China under Grants No.\ 11334014 and No.\ 61411136001.  
\end{acknowledgments}

\begin{widetext}

\begin{appendix}

\section{Solution of GIKE}
\label{app1}

We present the solution of GIKE in this part. The first order of GIKE [Eq.~(\ref{GE})] in clean limit reads
\begin{eqnarray}
&&\partial_t\delta\rho^{(1)}_{\bf
  k}\!+\!i[\xi_{\bf k}\!\tau_3\!+\!\Delta_{{\bf k}}\tau_1,\delta\rho^{(1)}_{\bf 
      k}]\!+\!i[\mu_{\rm
      eff}\tau_3\!+\!\delta\Delta^{(1)}_{\bf k}\tau_1,\rho^{(0)}_{\bf 
      k}]\!-\!\frac{i}{8m}[\tau_3,\nabla_{\bf R}^2\delta\rho_{\bf k}^{(1)}]\!+\!\frac{1}{2}\{{\bf v_k}\tau_3,{\bm \nabla}_{\bf R}\delta\rho^{(1)}_{\bf k}\}\!+\!\frac{1}{2}\{{\bf v_k}\tau_3,{\bm \nabla}_{\bf R}\rho^{(0)}_{\bf k}\}
  \nonumber\\
  &&\mbox{}\!+\!\frac{1}{2}\{e{\bf
    E}\tau_3\!-\!2i{\bf p}_s\tau_3\Delta_{\bf k}\tau_1,\partial_{\bf k}\rho^{(0)}_{\bf k}\}\!-\!\frac{i}{8}[{\bm \nabla}_{\bf R}{\bm \nabla}_{\bf R}\delta\Delta^{(1)}_{\bf
      k}\tau_1\!+\!2i{\bm \nabla}_{\bf R}{\bf p}_s\tau_3\Delta_{{\bf
        k}}\tau_1,\partial_{\bf k}\partial_{\bf k}\rho^{(0)}_{\bf
      k}]\!-\!\frac{1}{4m}[\nabla_{\bf R}\!\cdot{\bf p}_s\tau_3,\tau_3\rho_{\bf k}^{(0)}]=0.\label{FO}
\end{eqnarray}
In center-of-mass frequency and momentum space $[R=(t,{\bf R})\rightarrow{q=(\omega,{\bf q})}]$, by keeping the lowest three orders of ${\bf q}$, the components of above equation are written as
\begin{eqnarray}
&&i\omega\delta\rho^{(1)}_{{\bf k}0}={i{\bf v_k}\!\cdot\!{\bf q}}\delta\rho^{(1)}_{{\bf
      k}3}-e{\bf E}\!\cdot\!\partial_{\bf k}\rho^{(0)}_{{\bf k}3},\label{11}\\
&&i\omega\delta\rho^{(1)}_{{\bf k}3}=2\Delta_{{\bf k}}\delta\rho^{(1)}_{{\bf
      k}2}+{i{\bf v_k}\!\cdot\!{\bf q}}\delta\rho^{(1)}_{{\bf
      k}0}+{i}{\bf q}{\bf p}_{s}\Delta_{{\bf
      k}}:\partial_{\bf k}\partial_{\bf k}\rho^{(0)}_{{\bf
      k}1}/2+({\bf v_k}\cdot{i{\bf q}})({\bf v_k}\cdot{\bf p}_s)L_{\bf k},\label{12}\\
&&i\omega\delta\rho^{(1)}_{{\bf k}1}=-2\xi_{\bf k}\delta\rho_{{\bf k}2}^{(1)}-{i}{\bf q}{\bf p}_{s}\Delta_{{\bf
      k}}:\partial_{\bf k}\partial_{\bf k}\rho^{(0)}_{{\bf
      k}3}/2-{i{\bf q}\!\cdot\!{\bf p}_{s}}\rho^{(0)}_{{\bf k}1}/{(2m)},\label{13}\\
  &&i\omega\delta\rho^{(1)}_{{\bf k}2}=-2\Delta_{{\bf
      k}}\delta\rho_{{\bf k}3}^{(1)}+2\xi_{\bf k}\delta\rho_{{\bf
      k}1}^{(1)}-2\rho^{(0)}_{{\bf k}3}\delta\Delta^{(1)}_{\bf k}+2\rho^{(0)}_{{\bf
      k}1}\mu_{\rm eff}-\delta\Delta_{\bf k}^{(1)}({\bf q}\cdot\partial_{\bf k})^2\rho_{{\bf k}3}^{(0)}/4,\label{14}
\end{eqnarray}
where $L_{\bf k}=\partial_{{\bf v_k}\cdot{\bf p}_s}\rho^{(0)}_{{\bf k}0}$. Substituting Eqs.~(\ref{12}) and (\ref{13}) into Eq.~(\ref{14}), one has
\begin{eqnarray}
&&(i\omega)^2\delta\rho_{{\bf k}2}+2\Delta_{\bf k}[2\Delta_{{\bf k}}\delta\rho^{(1)}_{{\bf
      k}2}+{i{\bf v_k}\!\cdot\!{\bf q}}\delta\rho^{(1)}_{{\bf
      k}0}+{i}{\bf q}{\bf p}_{s}\Delta_{{\bf
      k}}:\partial_{\bf k}\partial_{\bf k}\rho^{(0)}_{{\bf
        k}1}/2+({\bf v_k}\cdot{i{\bf q}})({\bf v_k}\cdot{\bf p}_s)L_{\bf k}]+2\xi_{\bf k}[2\xi_{\bf k}\delta\rho_{{\bf k}2}^{(1)}\nonumber\\
 &&\mbox{}+{i}{\bf q}{\bf p}_{s}\Delta_{{\bf
      k}}:\partial_{\bf k}\partial_{\bf k}\rho^{(0)}_{{\bf
      k}3}/2+{i{\bf q}\!\cdot\!{\bf p}_{s}}\rho^{(0)}_{{\bf k}1}/{(2m)}]=2i\omega\rho^{(0)}_{{\bf
      k}1}\mu_{\rm eff}-2i\omega\rho^{(0)}_{{\bf k}3}\delta\Delta^{(1)}_{\bf k}-i\omega\delta\Delta_{\bf k}^{(1)}({\bf q}\cdot\partial_{\bf k})^2\rho_{{\bf k}3}^{(0)}/4.  
\end{eqnarray}
Further substituting Eq.~(\ref{11}) into above equation to replace $\delta\rho_{{\bf k}0}^{(1)}$, with Eq.~(\ref{14}), one immediately obtains 
\begin{eqnarray}
\delta\rho_{{\bf k}2}^{(1)}&=&\frac{1}{4E^2_{\bf k}\!-\!\omega^2}\Big\{2i\omega\rho^{(0)}_{{\bf
      k}1}\mu_{\rm eff}\!-\!2i\omega\rho^{(0)}_{{\bf k}3}\delta\Delta^{(1)}_{\bf k}\!-\!\frac{i\omega}{4}\delta\Delta_{\bf k}^{(1)}({\bf q}\cdot\partial_{\bf k})^2\rho_{{\bf k}3}^{(0)}\!-\!i\Delta_{\bf k}\xi_{\bf k}{\bf q}{\bf p}_s:{\partial_{\bf k}}{\partial_{\bf k}}\rho^{(0)}_{{\bf k}3}\!-\!i\Delta_{\bf k}^2{\bf q}{\bf p}_s:{\partial_{\bf k}}{\partial_{\bf k}}\rho^{(0)}_{{\bf k}1}  \nonumber\\
&&\mbox{}-\!\frac{i{\bf q}\cdot{\bf p}_s}{m}\xi_{\bf k}\rho^{(0)}_{{\bf k}1}-2i({\bf q}\!\cdot\!{\bf v_k})({\bf v}_{\bf k}\!\cdot\!{\bf p}_s)\Delta_{\bf k}L_{\bf k}\!+\!\frac{2\Delta_{\bf k}}{i\omega}(i{\bf q}\!\cdot\!{\bf v_k})(e{\bf E}\!\cdot\!{\partial_{\bf k}})\rho^{(0)}_{{\bf k}3}\!-\!\Big(\frac{{\bf v_k}\cdot{\bf q}}{\omega}\Big)^2[(\omega^2\!-\!4\xi_{\bf k}^2)\delta\rho_{{\bf k}2}^{(1)}\nonumber\\
&&\mbox{}+\!2i\omega\rho^{(0)}_{{\bf
      k}1}\mu_{\rm eff}\!-\!2i\omega\rho^{(0)}_{{\bf k}3}\delta\Delta^{(1)}_{\bf k}]\}.   
\end{eqnarray}
Then, after the first-order iteration, one obtains the solution of  $\delta\rho_{{\bf k}2}^{(1)}$:
\begin{eqnarray}
\delta\rho_{{\bf k}2}^{(1)}&=&\frac{1}{4E^2_{\bf k}\!-\!\omega^2}\Big\{2i\omega\rho^{(0)}_{{\bf
      k}1}\mu_{\rm eff}\!-\!2i\omega\rho^{(0)}_{{\bf k}3}\delta\Delta^{(1)}_{\bf k}\!-\!\frac{i\omega}{4}\delta\Delta_{\bf k}^{(1)}({\bf q}\cdot\partial_{\bf k})^2\rho_{{\bf k}3}^{(0)}\!-\!i\Delta_{\bf k}\xi_{\bf k}{\bf q}{\bf p}_s:{\partial_{\bf k}}{\partial_{\bf k}}\rho^{(0)}_{{\bf k}3}\!-\!i\Delta_{\bf k}^2{\bf q}{\bf p}_s:{\partial_{\bf k}}{\partial_{\bf k}}\rho^{(0)}_{{\bf k}1}  \nonumber\\
&&\mbox{}-\!\frac{i{\bf q}\cdot{\bf p}_s}{m}\xi_{\bf k}\rho^{(0)}_{{\bf k}1}\!-\!2i({\bf q}\!\cdot\!{\bf v_k})({\bf v}_{\bf k}\!\cdot\!{\bf p}_s)\Delta_{\bf k}L_{\bf k}\!+\!\frac{2\Delta_{\bf k}}{i\omega}(i{\bf q}\cdot{\bf v_k})(e{\bf E}\cdot{\partial_{\bf k}})\rho^{(0)}_{{\bf k}3}\!-\!\Big(\frac{{\bf v_k}\cdot{\bf q}}{\omega}\Big)^24\Delta_{\bf k}^2/(4E^2_{\bf k}\!-\!\omega^2)\nonumber\\
&&\mbox{}\times[2i\omega\rho^{(0)}_{{\bf
      k}1}\mu_{\rm eff}\!-\!2i\omega\rho^{(0)}_{{\bf k}3}\delta\Delta^{(1)}_{\bf k}]\Big\},\label{r12}
\end{eqnarray}
and hence, the solution of $\delta\rho_{{\bf k}1}^{(1)}$ from Eq.~(\ref{13}) is given by 
\begin{eqnarray}
\delta\rho_{{\bf k}1}^{(1)}&=&\frac{-2\xi_{\bf k}/(i\omega)}{4E^2_{\bf k}\!-\!\omega^2}\Big\{2i\omega\rho^{(0)}_{{\bf
      k}1}\mu_{\rm eff}\!-\!2i\omega\rho^{(0)}_{{\bf k}3}\delta\Delta^{(1)}_{\bf k}\!-\!\frac{i\omega}{4}\delta\Delta_{\bf k}^{(1)}({\bf q}\cdot\partial_{\bf k})^2\rho_{{\bf k}3}^{(0)}\!-\!i\Delta_{\bf k}\xi_{\bf k}{\bf q}{\bf p}_s:{\partial_{\bf k}}{\partial_{\bf k}}\rho^{(0)}_{{\bf k}3}\!-\!i\Delta_{\bf k}^2{\bf q}{\bf p}_s:{\partial_{\bf k}}{\partial_{\bf k}}\rho^{(0)}_{{\bf k}1}  \nonumber\\
&&\mbox{}-\!\frac{i{\bf q}\cdot{\bf p}_s}{m}\xi_{\bf k}\rho^{(0)}_{{\bf k}1}\!-\!2i({\bf q}\!\cdot\!{\bf v_k})({\bf v}_{\bf k}\!\cdot\!{\bf p}_s)\Delta_{\bf k}L_{\bf k}\!+\!\frac{2\Delta_{\bf k}}{i\omega}(i{\bf q}\cdot{\bf v_k})(e{\bf E}\cdot{\partial_{\bf k}})\rho^{(0)}_{{\bf k}3}\!-\!\Big(\frac{{\bf v_k}\cdot{\bf q}}{\omega}\Big)^24\Delta_{\bf k}^2/(4E^2_{\bf k}\!-\!\omega^2)\nonumber\\
&&\mbox{}\times[2i\omega\rho^{(0)}_{{\bf
      k}1}\mu_{\rm eff}\!-\!2i\omega\rho^{(0)}_{{\bf k}3}\delta\Delta^{(1)}_{\bf k}]\Big\}-\frac{{i}{\bf q}{\bf p}_{s}}{2i\omega}\Delta_{{\bf
      k}}:\partial_{\bf k}\partial_{\bf k}\rho^{(0)}_{{\bf
      k}3}-\frac{{i{\bf q}\!\cdot\!{\bf p}_{s}}}{2im\omega}\rho^{(0)}_{{\bf k}1}.\label{r11}
\end{eqnarray}
Moreover, for density-related $\delta\rho_{{\bf k}3}^{(1)}$ [Eq.~(\ref{density})] and current-related $\delta\rho_{{\bf k}0}^{(1)}$ [Eq.~(\ref{current})], substituting the solved $\delta\rho_{{\bf k}2}^{(1)}$ and $\delta\rho_{{\bf k}1}^{(1)}$ into Eq.~(\ref{14}) and keeping the lowest order of ${\bf q}$, one obtains the solution of $\delta\rho_{{\bf k}3}^{(1)}$:
\begin{equation}
\delta\rho_{{\bf k}3}^{(1)}
=[(\omega^2-4\xi^2_{\bf k})/(4E_{\bf k}^2-\omega^2)+1](2\rho^{(0)}_{{\bf
      k}1}\mu_{\rm eff}\!-\!2\rho^{(0)}_{{\bf k}3}\delta\Delta^{(1)}_{\bf k})/({2\Delta_{\bf k}})=4\Delta_{\bf k}/(4E_{\bf k}^2-\omega^2)(\rho^{(0)}_{{\bf
      k}1}\mu_{\rm eff}\!-\!\rho^{(0)}_{{\bf k}3}\delta\Delta^{(1)}_{\bf k}),\label{r13}
\end{equation}
and then, the solution of $\delta\rho_{{\bf k}0}^{(1)}$ from Eq.~(\ref{11}) reads
\begin{equation}
i\omega\delta\rho^{(1)}_{{\bf k}0}={i{\bf v_k}\!\cdot\!{\bf q}}4\Delta_{\bf k}/(4E_{\bf k}^2-\omega^2)(\rho^{(0)}_{{\bf
      k}1}\mu_{\rm eff}\!-\!\rho^{(0)}_{{\bf k}3}\delta\Delta^{(1)}_{\bf k})-e{\bf E}\!\cdot\!\partial_{\bf k}\rho^{(0)}_{{\bf k}3}.\label{r10}
\end{equation}

The second-order GIKE [Eq.~(\ref{GE})]  in center-of-mass frequency and momentum space is written as
\begin{eqnarray}
2i\omega\delta\rho_{\bf k}^{(2)}+i[\xi_{\bf k}\tau_3+\Delta_{\bf k}\tau_1,\delta\rho_{\bf k}^{(2)}]+i\Big[\frac{p_s^2}{2m}\tau_3+\delta\Delta_{\bf k}^{(2)}\tau_1,\rho_{\bf k}^{(0)}\Big]+\frac{1}{2}\{e{\bf E}\tau_3-2i{\bf p}_s\tau_3\Delta_{\bf k}\tau_1\}+\frac{i}{2}[{\bf p}_s{\bf p}_s\Delta_{\bf k}\tau_1,\partial_{\bf k}\partial_{\bf k}\rho_{\bf k}^{(0)}]=0, \nonumber\\ 
\end{eqnarray}
in which we only keep the lowest order of ${\bf q}$ to consider a homogeneous influence/excitation from ${\bf p}_s$. The components of above equation are written as
\begin{eqnarray}
  &&2i\omega\delta\rho^{(2)}_{{\bf k}0}=-e{\bf E}\cdot\partial_{\bf k}\delta\rho^{(1)}_{{\bf k}3},\label{21}\\
&&2i\omega\delta\rho^{(2)}_{{\bf k}3}=2\Delta_{{\bf k}}\delta\rho^{(2)}_{{\bf
      k}2}-e{\bf E}\cdot{\partial_{\bf k}}\delta\rho^{(1)}_{{\bf k}0},\label{22}\\
&&2i\omega\delta\rho^{(2)}_{{\bf k}1}=-2\xi_{\bf k}\delta\rho_{{\bf k}2}^{(2)},\label{23}\\
  &&2i\omega\delta\rho^{(2)}_{{\bf k}2}=-2\Delta_{{\bf
      k}}\delta\rho_{{\bf k}3}^{(2)}+2\xi_{\bf k}\delta\rho_{{\bf
      k}1}^{(2)}-2\rho^{(0)}_{{\bf k}3}\delta\Delta^{(2)}_{\bf k}+2\rho^{(0)}_{{\bf
      k}1}p_s^2/(2m)-2\Delta_{\bf k}{\bf p}_s\cdot{\partial_{\bf k}}\delta\rho^{(1)}_{{\bf k}0}-\Delta_{\bf k}({\bf p}_s\cdot\partial_{\bf k})^2\rho_{{\bf k}3}^{(0)}.\label{24}
\end{eqnarray}
Carrying Eqs.~(\ref{22})-(\ref{23}) and Eq.~(\ref{11}) into Eq.~(\ref{24}), one has
\begin{equation}
\delta\rho^{(2)}_{{\bf k}2}=\frac{1}{4E_{\bf k}^2-4\omega^2}\Big\{2i\omega[2\rho^{(0)}_{{\bf
      k}1}\mu^{(2)}_{\rm eff}-2\rho^{(0)}_{{\bf k}3}\delta\Delta^{(2)}_{\bf k}]+2i\omega\Delta_{\bf k}\Big[\Big(\frac{e{\bf E}}{i\omega}-{\bf p}_s\Big)\cdot{\partial_{\bf k}}\Big]^2\rho_{{\bf k}3}^{(0)}\Big\},\label{r22}
\end{equation}
and then, from Eq.~(\ref{23}), one obtains
\begin{equation}
\delta\rho^{(2)}_{{\bf k}1}=-\frac{\xi_{\bf k}/(i\omega)}{4E_{\bf k}^2-4\omega^2}\Big\{2i\omega[2\rho^{(0)}_{{\bf
      k}1}\mu^{(2)}_{\rm eff}-2\rho^{(0)}_{{\bf k}3}\delta\Delta^{(2)}_{\bf k}]+2i\omega\Delta_{\bf k}\Big[\Big(\frac{e{\bf E}}{i\omega}-{\bf p}_s\Big)\cdot{\partial_{\bf k}}\Big]^2\rho_{{\bf k}3}^{(0)}\Big\}.\label{r21}
\end{equation}

\section{Phase and gap equations as well as superfluid density}
\label{app2}

In this part, we present the derivation of the phase and gap equations as well as superfluid density. We first solve the Hartree field. Substituting $\rho_{{\bf k}3}^{(0)}$ [Eq.~(\ref{r0})] and the solved $\delta\rho_{{\bf k}3}^{(1)}$ [Eq.~(\ref{r13})] and Eq.~(\ref{22}) into Eq.~(\ref{density}), at low-frequency regime, one has
\begin{equation}\label{n11}
\delta{n}=2\sum_{\bf k}\Big(\frac{\xi_{\bf k}}{E_{\bf k}}\mu^{(1)}_{\rm eff}\partial_{\xi_{\bf k}}\rho_{{\bf k}3}^{q0}+\frac{\Delta_{\bf k}}{E_{\bf k}^2}\rho_{{\bf k}1}^{(0)}\mu^{(1)}_{\rm eff}\Big)+\frac{2}{i\omega}{\sum_{\bf k}}'\Delta_{\bf k}\delta\rho_{{\bf k}2}^{(2)}/(i\omega),
\end{equation}
where $\rho_{{\bf k}3}^{q0}=[{f(E^+_{\bf k})-f(E^-_{\bf k})}]/{2}$ represents the $\tau_3$ component of the equilibrium density of matrix in quasiparticle space. On the right-hand side of above equation, substituting Eq.~(\ref{gap}) to replace $\Delta_{\bf k}$ in the last term, with Eq.~(\ref{phase}), one has 
\begin{eqnarray}\label{n12}
  \delta{n}&=&2\sum_{\bf k}\Big[\frac{\xi_{\bf k}}{E_{\bf k}}\mu^{(1)}_{\rm eff}\partial_{\xi_{\bf k}}\rho_{{\bf k}3}^{q0}+\frac{\Delta_{\bf k}}{E_{\bf k}^2}\rho_{{\bf k}1}^{(0)}\mu^{(1)}_{\rm eff}\Big]-\frac{2}{i\omega}{\sum_{{\bf k}'{\bf k}}}'g_{\bf kk'}\rho_{{\bf k'}1}\delta\rho_{{\bf k}2}^{(2)}\nonumber\\
  &=&2\sum_{\bf k}\Big[\frac{\xi_{\bf k}}{E_{\bf k}}\mu^{(1)}_{\rm eff}\partial_{\xi_{\bf k}}\rho_{{\bf k}3}^{q0}+\frac{\Delta_{\bf k}}{E_{\bf k}^2}\rho_{{\bf k}1}^{(0)}\mu^{(1)}_{\rm eff}\Big]-\frac{2}{i\omega}{\sum_{{\bf k}'}}'\rho_{{\bf k'}1}\Big[{\sum_{\bf k}}'g_{\bf k'k}\delta\rho_{{\bf k}2}^{(2)}\Big]\nonumber\\
&=&2\sum_{\bf k}\partial_{\xi_{\bf k}}(\xi_{\bf k}F_{\bf k})\mu^{(1)}_{\rm eff}=-2D\mu^{(1)}_{\rm eff}.
\end{eqnarray}
Consequently, Eq.~(\ref{density1}) and Hartree field $\mu_H=-2V_{q}D\mu_{\rm eff}=-2DV_q(i\omega\delta\theta/2+U)/(1+2DV_q)$ are obtained. Then, the electric field reads
\begin{equation}
e{\bf E}=i{\bf q}(\mu_H+U)=\frac{i{\bf q}(U-2DV_qi\omega\theta/2)}{1+2DV_q}\approx{i\omega}{\bf p}_s,\label{see} 
\end{equation}
in which we have taken the long-wave approximation ($DV_q\gg1$).

Substituting the solved $\delta\rho_{{\bf k}2}^{(1)}$ [Eq.~(\ref{r12})] and electric field [Eq.~(\ref{see})] into Eq.~(\ref{phase}), at low-frequency regime, one has
\begin{equation}
{\sum_{\bf k}}'g_{\bf k'k}\Big\{2i\omega\mu_{\rm eff}\frac{\rho_{{\bf k}1}^{(0)}}{4E_{\bf k}^2}\!+\!\frac{2({\bf v_k}\!\cdot\!{i{\bf q}})({\bf v_k}\!\cdot\!{\bf p}_s)}{4E_{\bf k}^2}\Big[\frac{\Delta_{\bf k}^3}{E_{\bf k}^3}\rho^{q0}_{{\bf k}3}\!-\!\Delta_{\bf k}L_{\bf k}\!+\!2\Delta_{\bf k}\Big(\frac{\xi_{\bf k}}{E_{\bf k}}\Big)^2\partial_{E_{\bf k}}\rho^{q0}_{{\bf k}3}\!-\!\frac{1}{2}(\Delta_{\bf k}^2\partial^2_{\xi_{\bf k}}\rho_{{\bf k}1}^{(0)}\!+\!\Delta_{\bf k}\xi_{\bf k}\partial^2_{\xi_{\bf k}}\rho_{{\bf k}3}^{(0)})\Big]\Big\}=0,
\end{equation}
where we have taken care of the particle-hole symmetry to remove terms with the odd order of $\xi_{\bf k}$ in the summation of ${\bf k}$. Taking a generalized pairing potential $g_{\bf kk'}=g\cos(\zeta\theta_{\bf k}+\alpha)\cos(\zeta\theta_{\bf k'}+\alpha)$ which gives rise to {\small $\Delta_{{\bf k}}=\Delta\cos(\zeta\theta_{\bf k}+\alpha)$}, the above equation becomes
\begin{equation}
  2i\omega\mu_{\rm eff}Z_1+{i{\bf q}\cdot{\bf p}_s}Z_2/m=0,
\end{equation}
with $Z_1={\sum_{\bf k}}'\frac{\Delta_{\bf k}^2}{E_{\bf k}^3}\rho_{{\bf k}3}^{q0}\approx{\sum_{\bf k}}'\partial_{\xi_{\bf k}}(\xi_{\bf k}F_{\bf k})\approx{-D}$ around Fermi surface and 
\begin{equation}
  Z_2=\!\frac{k_F^2}{m}{\sum_{\bf k}}'\frac{\Delta_{\bf k}^2}{E_{\bf k}^2}\Big[\frac{\Delta_{\bf k}^2}{E_{\bf k}^3}\rho_{{\bf k}3}^{q0}\!-\!L_{\bf k}\!+\!\frac{\xi_{\bf k}^2}{E_{\bf k}^2}\partial_{E_{\bf k}}\rho_{{\bf k}3}^{q0}\!-\!\frac{1}{2}(\Delta_{\bf k}\partial^2_{\xi_{\bf k}}\rho_{{\bf k}1}^{(0)}\!+\!\xi_{\bf k}\partial^2_{\xi_{\bf k}}\rho_{{\bf k}3}^{(0)})\Big]\!=\!\frac{k_F^2}{m}{\sum_{\bf k}}'\frac{\Delta_{\bf k}^2}{E_{\bf k}^2}\Big(\frac{\Delta_{\bf k}^2}{E_{\bf k}^3}\!+\!\frac{\xi_{\bf k}^2}{E_{\bf k}^3}\!-\!\partial_{E_{\bf k}}\Big)\rho_{{\bf k}3}^{q0}\!=\!-n_s.
\end{equation}
Then, Eq.~(\ref{PE1}) is obtained. 

Substituting the solved $\delta\rho_{{\bf k}1}^{(1)}$ [Eq.~(\ref{r11})] into Eq.~(\ref{gap}) and taking care of the particle-hole symmetry to remove terms with the odd order of $\xi_{\bf k}$ in the summation of ${\bf k}$, at low-frequency regime, one has
\begin{eqnarray}
  &&\delta\Delta_{\bf k'}^{(1)}+\sum_{\bf k}g_{\bf k'k}\delta\Delta_{\bf k}^{(1)}\frac{\xi_{\bf k}}{2E^2_{\bf k}}\Big[2\rho^{(0)}_{{\bf k}3}\!+\!\Big(\frac{{\bf q}\cdot\partial_{\bf k}}{2}\Big)^2\rho_{{\bf k}3}^{(0)}\!-\!2\Big(\frac{{\bf v_k}\cdot{\bf q}}{\omega}\Big)^2\frac{\Delta_{\bf k}^2\rho^{(0)}_{{\bf k}3}}{E^2_{\bf k}}\Big]=\frac{i{\bf q}\cdot{\bf p}_s}{2im\omega}\sum_{\bf k}g_{\bf k'k}\Big\{\frac{-\xi_{\bf k}}{E^2_{\bf k}}\Big[\Delta_{\bf k}\xi_{\bf k}\partial_{\xi_{\bf k}}\rho^{(0)}_{{\bf k}3}\nonumber\\
&&\mbox{}\!+\!\Delta_{\bf k}^2\partial_{\xi_{\bf k}}\rho^{(0)}_{{\bf k}1}\!+\!\xi_{\bf k}\rho^{(0)}_{{\bf k}1}\Big]+\Delta_{{\bf
      k}}\partial_{\xi_{\bf k}}\rho^{(0)}_{{\bf
      k}3}+\rho^{(0)}_{{\bf k}1}\Big\}.
\end{eqnarray}
Then, consider a homogeneous influence/excitation from ${\bf p}_s$, one finds a vanishing $\delta\Delta_{\bf k}^{(1)}$. Moreover, with Eq.~(\ref{see}), substituting the solved $\delta\rho_{{\bf k}1}^{(2)}$ [Eq.~(\ref{r21})] into Eq.~(\ref{gap}) and taking care of the particle-hole symmetry to remove terms with the odd order of $\xi_{\bf k}$ in the summation of ${\bf k}$, one immediately finds $\delta\Delta_{\bf k}^{(2)}=0$.  

Furthermore, with $\delta\Delta_{\bf k}^{(1)}=0$ and the solved Hartree field as well as Eq.~(\ref{see}), Eq.~(\ref{r10}) at low-frequency regime becomes:
\begin{equation}
  \delta\rho^{(1)}_{{\bf k}0}=-{{\bf v_k}\!\cdot\!{\bf p}_s}\partial_{\bf k}\rho^{(0)}_{{\bf k}3},
\end{equation}
in which we have taken the long-wave approximation ($DV_q\gg1$). Consequently, the current-related $\rho_{{\bf k}0}$ [Eq.~(\ref{current})] reads
\begin{equation}
  \rho_{{\bf k}0}=\rho^{(0)}_{{\bf k}0}+\delta\rho^{(1)}_{{\bf k}0}=({{\bf v_k}\!\cdot\!{\bf p}_s})(\partial_{E_{\bf k}}\rho_{{\bf k}3}^{q0}-\partial_{\bf k}\rho^{(0)}_{{\bf k}3})=({{\bf v_k}\!\cdot\!{\bf p}_s})\frac{\Delta^2_{\bf k}}{E_{\bf k}^2}\Big(\partial_{E_{\bf k}}-\frac{1}{E_{\bf k}}\Big)\rho_{{\bf k}3}^{q0}=({{\bf v_k}\!\cdot\!{\bf p}_s})\frac{\Delta^2_{\bf k}}{E_{\bf k}}\partial_{E_{\bf k}}F_{\bf k}.
\end{equation}
Then, Eq.~(\ref{rk0}) is obtained.

\section{Derivation of Eq.~(\ref{GFMM})}
\label{app3}

In this part, we present the derivation of Eq.~(\ref{GFMM}). With the action in Eq.~(\ref{action}), we first calculate the quantum fluctuation part in Eq.~(\ref{GFM}):
\begin{eqnarray}
  (p_s^{\phi})^2|_{\rm Q}\!\!&\!=\!&\!\!\sum_qq^2\Big[\Big\langle\Big|\frac{\delta\theta^*_{q{\bf e}_{\phi}}(t+0^+)}{2}\frac{\delta\theta_{q{\bf e}_{\phi}}(t)}{2}e^{iS}\Big|\Big\rangle\!=\!\!\!\sum_qq^2\frac{1}{Z_0}{\int}D\theta{D\theta^*}\frac{\delta\theta^*_{q{\bf e}_{\phi}}(t+0^+)}{2}\frac{\delta\theta_{q{\bf e}_{\phi}}(t)}{2}e^{iS} \nonumber\\
    &=&\!\!\!\sum_qi^2q^2\frac{1}{Z_0}{\int}D\theta{D\theta^*}\delta_{J_{q^*{\bf e}_{\phi}}(t+0^+)}\delta_{J_{q{\bf e}_{\phi}}(t)}e^{i[S+\int{dt'}\sum_{\bf q'}(J_{\bf q'}\theta_{\bf q'}/2+J_{\bf q'}^*\theta^*_{\bf q'}/2)]}\Big|_{J=J^*=0}\nonumber\\
    &=&\!\!\!\sum_qi^2q^2\delta_{J_{q^*{\bf e}_{\phi}}(t+0^+)}\delta_{J_{q{\bf e}_{\phi}}(t)}\exp\Big\{i\int{dt'}\sum_{\bf q'}(J_{\bf q'}+D_{q'}U_{\bf q'}^*\partial_{t'})D(t',{\bf q}')(J^*_{\bf q'}+D_{q'}U_{\bf q'}\partial_{t'})\Big\}\Big|_{J=J^*=0}  \nonumber\\
    &=&\!\!\!\sum_qq^2{\rm Tr}_{\omega}[D^2(t,q{\bf e}_{\phi})|D_qU_{q{\bf e}_{\phi}}\partial_t|^2]=\frac{i}{C}\sum_{q\omega}\frac{q^2\omega^2U_{q{\bf e}_{\phi}}U_{-q{\bf e}_{\phi}}}{(\omega^2\!-\!\omega_N^2+i0^+)^2}.\label{psq}
\end{eqnarray}
Here, the Green function $D(t,{\bf q})=D_q^{-1}(\partial_t^2+\omega_N^2)^{-1}$; $J_{\bf q}$ denotes the generating functional and $\delta_{J_{\bf q}}$ represents the functional derivative\cite{FT}. 

We next derive the thermal fluctuation part in Eq.~(\ref{GFM}). By mapping the action in Eq.~(\ref{action}) into the imaginary-time one $\mathcal{S}$:
\begin{equation}
\mathcal{S}=\int^{\beta}_{0}d\tau\sum_{\bf q}D_q\Big[{\theta_{\bf q}^*}(\partial_{\tau}^2-\omega_N^2){\theta_{\bf q}}/4-iU_{\bf q}^*\partial_{\tau}\theta_{\bf q}/2+iU_{\bf q}\partial_{\tau}\theta_{\bf q}^*/2\Big].  
\end{equation}
Since the Bosonic field $\theta(\tau=\beta)=\theta(\tau=0)$\cite{G1}, the Josephson coupling terms (the last two terms) vanish in 
the imaginary-time space, i.e., the Josephson effect from electric potential, as a quantum effect, can not excite the imaginary-time fluctuation of the superconducting phase. The thermal fluctuation part in Eq.~(\ref{GFM}) then reads
\begin{eqnarray}
  (p_s^{\phi})^2|_{\rm T}\!\!&\!=\!&\!\!\sum_qq^2\Big[\Big\langle\Big|\frac{\delta\theta^*_{q{\bf e}_{\phi}}(\tau)}{2}\frac{\delta\theta_{q{\bf e}_{\phi}}(\tau)}{2}e^{-\mathcal{S}}\Big|\Big\rangle\Big]\!=\!\!\!\sum_qq^2\Big[\frac{1}{\mathcal{Z}_0}{\int}D\theta{D\theta^*}\frac{\delta\theta^*_{q{\bf e}_{\phi}}(\tau)}{2}\frac{\delta\theta_{q{\bf e}_{\phi}}(\tau)}{2}e^{-\mathcal{S}}\Big]\nonumber\\
   &=&\!\!\!\sum_qq^2\frac{1}{\mathcal{Z}_0}{\int}D\theta{D\theta^*}\delta_{J_{q^*{\bf e}_{\phi}}}\delta_{J_{q{\bf e}_{\phi}}}e^{-[\mathcal{S}+\int{d\tau}\sum_{\bf q'}(J_{\bf q'}\theta_{\bf q'}/2+J_{\bf q'}^*\theta^*_{\bf q'}/2)]}\Big|_{J=J^*=0}\nonumber\\
    &=&\!\!\!\sum_qq^2\delta_{J_{q^*{\bf e}_{\phi}}}\delta_{J_{q{\bf e}_{\phi}}}\exp\Big\{-\int{dt'}\sum_{\bf q'}J_{\bf q'}D(\tau,{\bf q}')J^*_{\bf q'}\Big\}\Big|_{J=J^*=0}  \nonumber\\
    &=&\!\!-\!\sum_qq^2{\rm Tr}_{\omega}[D(\tau,q{\bf e}_{\phi})]=\!-\!\frac{1}{\beta}\sum_{q\omega_n}\frac{q^2}{D_q}\frac{1}{(i\omega_n)^2\!-\!\omega_N^2}. \label{pst}
\end{eqnarray}
Consequently, with Eqs.~(\ref{psq}) and (\ref{pst}), Eq.~(\ref{GFMM}) is derived.

\section{Derivation of Eqs.~(\ref{Q1}) and (\ref{NDE})}
\label{app4}

In this part, we present the derivation of Eqs.~(\ref{Q1}) and (\ref{NDE}). Following the standard Matsubara-frequency summation\cite{G1}, from Eq.~(\ref{GFP}) one has
\begin{eqnarray}
  Q_1&=&V_f+\sum_p{\rm Tr}[G_0(p)\tau_3]=V_f+\sum_{\bf k}\frac{1}{\beta}\sum_{ip_n}\frac{2\xi_{\bf k}}{(ip_n-E_{\bf k}^+)(ip_n-E_{\bf k}^-)}=V_f+\sum_{\bf k}\frac{2\xi_{\bf k}[f(E_{\bf k}^+)-(E_{\bf k}^-)]}{2E_{\bf k}} \nonumber\\
  &=&\sum_{\bf k}(1+2\xi_{\bf k}F_{\bf k})=-\frac{k_F^2}{m}{\sum_{\bf k}}'\partial_{\xi_{\bf k}}\big(\xi_{\bf k}F_{\bf k}\big), 
\end{eqnarray}
and
\begin{eqnarray}
 && Q_2=-\frac{1}{2}\sum_p{\rm Tr}[G_0(p)\tau_3G_0(p)\tau_3]=-\sum_{\bf k}\frac{1}{\beta}\sum_{ip_n}\frac{(ip_n-{\bf p}_s\cdot{\bf v_k})^2+\xi_{\bf k}^2-\Delta_{\bf k}^2}{(ip_n-E_{\bf k}^+)^2(ip_n-E_{\bf k}^-)^2}=-\sum_{\bf k}\sum_{\eta=\pm}\Big[\frac{E_{\bf k}^2+\xi_{\bf k}^2-\Delta_{\bf k}^2}{4E_{\bf k}^2}\partial_{E_{\bf k}^{\eta}}f(E_{\bf k}^{\eta})\nonumber\\
  &&\mbox{}+\eta\Big(\frac{2E_{\bf k}}{4E_{\bf k}^2}-\frac{E_{\bf k}^2+\xi_{\bf k}^2-\Delta_{\bf k}^2}{4E_{\bf k}^3}\Big)f(E_{\bf k}^{\eta})\Big]=-{\sum_{\bf k}}\Big\{\frac{\Delta_{\bf k}^2}{E_{\bf k}^2}F_{\bf k}+\frac{\xi_{\bf k}^2}{E_{\bf k}^2}\partial_{E_{\bf k}}\Big[\frac{f(E_{\bf k}^+)-f(E_{\bf k}^-)}{2}\Big]\Big\}=-{\sum_{\bf k}}'\partial_{\xi_{\bf k}}\big(\xi_{\bf k}F_{\bf k}\big). 
\end{eqnarray}
Then, Eqs.~(\ref{Q1}) and (\ref{NDE}) are obtained. 

\end{appendix}

\end{widetext}


\begin{thebibliography}{0}
\bibitem{SNT} In the original phenomenological two-fluid model which was first proposed by Tisza and London\cite{Tisza} in superconductors and then developed by Landau\cite{Landau} in Bosonic liquid helium, the superfluid and normal fluid in superconductors describe the excitations in superconducting and normal states\cite{Ba0}, respectively. From our recent study in microscopic view\cite{GIKE1}, we followed the idea of Fulde-Ferrell-Larkin-Ovchinnikov state\cite{FF1,FF2} and then divided the BCS spherical shell in momentum space into the pairing and unpairing regions with finite and vanishing anomalous correlation, respectively. It was further shown that the particles in unpairing region behave like normal ones, whereas the dynamics in the pairing region consists of the contributions from Meissner density and Bogoliubov quasiparticles\cite{GIKE1}. We therefore considered the excitations in the pairing and unpairing regions as superfluid and normal fluid, and then, recovered the optical conductivity of the two-fluid model at low temperature\cite{GIKE1}.
\end{thebibliography}
\end{document}